\documentclass[hidelinks,onefignum,onetabnum]{siamart220329}
\usepackage{subcaption}
\usepackage{colortbl}
\usepackage{lipsum}
\usepackage{amsfonts}
\usepackage{graphicx}
\usepackage{algorithmic}

\usepackage{tabularx,booktabs}
\usepackage{multirow}

\usepackage{xcolor} 
\newcommand{\Order}[1]{\ensuremath{\mathcal{O}(#1)}}    % big O notation

\begin{document}

%\title{%
%A performance portable, fully implicit Landau collision operator with batched linear solvers
%\funding{This work was funded by the Fog Research Institute under contract no.~FRI-454.}} \\[2ex]
%{\normalfont \normalsize \sffamily  \textcolor{black}{``in fond memory of Ravindra Samtaney''}}
%}

\title{A performance portable, fully implicit Landau collision operator with batched linear solvers}

\author{Mark F. Adams\thanks{Lawrence Berkeley National Laboratory, Berkeley, CA (\email{mfadams@lbl.gov})}.
\and Peng Wang\thanks{{NVIDIA} Corporation, Santa Clara, CA}.
\and Jacob Merson\thanks{Rensselaer Polytechnic Institute, Troy, NY.}
\and Kevin Huck\thanks{University of Oregon, Eugene, OR.}
\and Matthew G. Knepley\thanks{University of Buffalo, Buffalo, NY.}
}

\maketitle 
\begin{center}
{\normalfont \normalsize \sffamily {\textit{In fond memory of Ravindra Samtaney}}} \\ [2ex]
\end{center}

\begin{abstract}
Modern accelerators use hierarchical parallel programming models that enable massive multithreading within a processing element (PE), with multiple PEs per device driven by traditional processes.
\textit{Batching} is a technique for exposing PE-level parallelism in algorithms that have traditionally run on MPI processes or multiple threads within a single process.
Opportunities for batching arise in, for example, kinetic discretizations of magnetized plasmas where collisions are advanced in velocity space at each spatial point independently.

This paper builds on previous work on a high-performance, fully nonlinear, Landau collision operator by batching the linear solver, as well as batching the spatial point problems and adding new support for multiple grids for multiscale, multi-species problems.
An anisotropic relaxation verification test that agrees well with previous published results and analytical models is presented.
The performance results from NVIDIA A100 and AMD MI250X nodes are presented with hardware utilization analysis for each architecture.
The entire implicit Landau operator time advance is implemented in Kokkos for performance portability, running entirely on the device and is available in the PETSc numerical library.

\end{abstract}

% REQUIRED
\begin{keywords}
Batch solvers, Landau collision operator, GPU finite elements, kinetic methods
\end{keywords}

% REQUIRED
\begin{MSCcodes}
76X05, 68N01, 65F10, 65F50, 65Y20, 65Z05
\end{MSCcodes}

\section{Introduction}

The programming model used for modern accelerator hardware, introduced in the \textit{CUDA} programming language, is now supported in several languages and libraries, such as Kokkos \cite{CarterEdwards20143202}.
This model supports massive multithreading within a processing element (PE) and encourages vector processing with C/C++ syntax, in addition to traditional MPI processes for coarse-grain parallelism.
The Kokkos programming model supports accelerator languages, such as \textit{SYCL}, \textit{HIP} and \textit{CUDA}, as well as \textit{OpenMP}.

High-dimensional applications with tensor product structures, such as combustion with chemistry \cite{AggarwalBatchChem}, kinetic methods with collision operators \cite{Hager2016,Hirvijoki2016}, ensemble problems in uncertainty quantification \cite{LIEGEOIS2020113188,Liegeois2023}, and others \cite{BOUKARAM201819,merson2023} run many small problems independently.
These problems have traditionally been run in MPI processes or with small-scale thread parallelism within a process.
However, these solves may be small enough to run effectively on a single PE.
Techniques known as \textit{batching} are designed to explicitly expose this PE-level parallelism to the accelerator.

Phase space problems, from $2D$ to $6D$ are used to accurately model many problems in computational physics.
Magnetized plasmas are one such application.
Plasmas are governed by the symplectic Vlasov-Maxwell system \cite{Vlasov1968}, coupled with a metric or dissipative collision operator.
This work focuses on Fokker-Planck collisions in Landau form, which is the gold standard for small-angle collision-dominated plasmas \cite{landau1936kinetic}, using discretizations that preserve the geometric structure of this system in the \textit{metriplectic} formalism \cite{Hirvijoki2016,Kraus2017}.
The evolution of the collision operator is computed at each spatial point independently and is well-suited to batch processing, including the linear solver in a nonlinear solver used in implicit time integrators.

This paper builds on previous work on a high-performance, portable, structure-preserving, grid-based, Landau collision operator with high-order accurate finite element discretizations and block-structured adaptive mesh refinement (AMR) \cite{Hirvijoki2016,AdamsHirvijokiKnepleyBrownIsaacMills2017,Adams2022a}. Background from previous work is presented in \S\ref{sec:landau} followed by new material:
\begin{itemize}
    \item multiple grids and batching of the collision operator (\S\ref{sec:batching_mg}), 
    \item performance with a ten species test on NVIDIA and AMD device nodes (\S\ref{sec:throughput_perf}),
    \item unstructured adaptive meshing for collision operators (\S\ref{sec:meshing}),
    \item an anisotropic relaxation verification test (\S\ref{sec:verify}), 
    \item and performance data on NVIDIA A100 and AMD MI250X nodes (\S\ref{sec:perfanisotropic}),
    \item MI250X and A100 hardware utilization analysis (\S\ref{sec:utilization}),
\end{itemize}
and \S\ref{sec:conc} concludes the paper.
All codes are available in PETSc. 
Data, plotting scripts and reproducibility instructions are publicly available (Appendix \ref{sec:ad}).

\section{Landau collisions for magnetized plasmas}
\label{sec:landau}

This section reviews previous work on structure preserving methods for the time evolution of the Landau collision operator \cite{Hirvijoki2016,AdamsHirvijokiKnepleyBrownIsaacMills2017,Adams2022a}.
The density of each species $\alpha$, $f_\alpha \left(t,\vec{v}\right)$, evolves, in velocity space $\vec{v}\in\mathbb{R}^3$, from collisional effects with each species $\beta$ according to
$\frac{df_\alpha}{dt}=\sum_{\beta} C\left[f_\alpha,f_\beta\right]_{\alpha\beta}$, with
\begin{equation}
\label{eq:landau1}
C\left[f_\alpha,f_\beta\right]_{\alpha\beta} = 
\nu_{\alpha\beta}\frac{m_0}{m_{\alpha}}\nabla \cdot\int \limits_{\bar\Omega} d{\bar{v}}\;\mathbf{U}(\vec{v},{\bar{v}})\cdot\left(\frac{m_0}{m_{\alpha}}\bar{f}_{\beta}\nabla f_{\alpha} - \frac{m_0}{m_{\beta}}f_{\alpha} \bar \nabla \bar{f}_{\beta}\right),
\end{equation}
 a collision frequency $\nu_{\alpha\beta}=e_{\alpha}^2e_{\beta}^2\ln\Lambda_{\alpha\beta}/8\pi m_0^2\varepsilon_0^2$, the Coulomb logarithm $\ln\Lambda_{\alpha\beta}$, an arbitrary reference mass $m_0$, the vacuum permittivity $\varepsilon_0$ and the effective charges $e_\alpha$ of each species.
$\mathbf{U}(\vec{v},{\bar{v}})$ is the Landau tensor \cite{landau1936kinetic,Hirvijoki2016}.
Overbar terms are evaluated on the grid for the domain $\bar\Omega$ of species $\beta$ and $\bar v \equiv \vec{\bar{v}}$ for clarity.

This inner integration over all domains, $\bar \Omega$, results in an $\Order{N^2}$ work complexity algorithm, where $N$ is the sum of the number of species times the number of integration points on each species' grid.
This coupling between species is only through the inner integral and results in the property that each species can use a separate grid, which is common in physics codes \cite{Hager2016}.

\subsubsection{Nondimensionalization and Landau structure}
\label{ssec:landau_nond}

The system (\ref{eq:landau1}) is nondimensionalized with a reference speed $v_0$, such as the thermal speed of electrons, to define a velocity coordinate ${\vec{x} = \vec{v} / v_0}$, not to be confused with the spatial coordinate.
The distribution function variable is normalized with $\tilde f_{\alpha} = f_{\alpha}v_0^3/n_0$ with a number density $n_0$.
The electrons are used to set $v_0$ and $m_0$ in this work.
Time and the collision frequency are nondimensionalized with
\begin{equation}
%\label{eq:ndvars}
t_0 = \frac{8\pi m_0^2\varepsilon_0^2v_0^3}{e^4\ln\left(\Lambda_{ee}\right)n_0}, \quad 
\tilde \nu_{\alpha\beta} = \frac{t_0n_0}{v_0^3} \nu_{\alpha\beta},
\end{equation}
with elementary charge $e$.
Further, $d \vec{x} = v_0^{-3} d \vec{v}$, $\mathbf{U}( \vec{x},{\bar{x}}) = v_0\mathbf{U}(\ \vec{v},{\bar{v}})$  and $\frac{\partial}{\partial  \vec{x}} = v_0\frac{\partial}{\partial  \vec{v} }$.
% Note, with this normalization $\tilde \nu_{ss}=1$ if the Coulomb logarithm for electron-electron collisions is used.
% Any species can be used for normalization. % and, likewise, any species may be used for normalization.
%For the further details, see \cite{Hirvijoki2016, Adams2022a}.

\subsection{Discretizations}
\label{sec:discrete}

The time evolution of the collisional dynamics of the density of each species $f\left(\vec{x}\right)$ with that of all species $\beta$, after nondimensionalization and assuming all species are identical for clarity, can be written as 
\begin{equation}
\label{eq:land2}
\frac{df}{dt} = \sum_{\beta} \nabla \cdot \int \limits_{\bar\Omega} d{\bar{x}}\;\mathbf{U}(\vec{x},{\bar{x}})\cdot\left(\bar{f_\beta}\nabla f - f \bar \nabla \bar{f_\beta}\right) = \left( \nabla \cdot \mathbf{D}\left(\vec{x} \right) \nabla - \nabla \cdot \mathbf{K}\left(\vec{x} \right) \right) f,
\end{equation}
with tensor $\mathbf{D}\left(\vec{x} \right) \equiv \sum_{\beta} \int \limits_{\bar\Omega} d{\bar{x}}\;\mathbf{U}(\vec{x},{\bar{x}})\cdot\bar{f_\beta}$ and vector $\mathbf{K}\left(\vec{x} \right) \equiv \sum_{\beta} \int \limits_{\bar\Omega} d{\bar{x}}\;\mathbf{U}(\vec{x},{\bar{x}})\cdot \bar \nabla \bar{f_\beta}$.

This system is evolved in time with a ``solver stack" of time integrators, nonlinear and linear algebraic solvers, and discretizations.
The fully implicit time integrators in this paper, backward Euler or a three stage adaptive Runge-Kutta method, are sufficient for energy conservation but do not result in an entropy stable, or monotonic in entropy, method.
A discrete gradient time integrator can be entropy stable \cite{Kraus2017} and is under development.

A finite element discretization of (\ref{eq:landau1}) that conserves moments up to the order of the polynomial that can be represented exactly by the finite element space was developed by Hirvijoki and Adams \cite{Hirvijoki2016}.
For example quadratic elements, $P2$ or $Q2$, conserves up to the second moment, which is energy and is of particular interest to physicists.
The spatial discretization enforces weak equivalence by multiplying (\ref{eq:land2}) with a test function $\phi_i$ in a set of basis vectors and integrating over the domain in a standard finite element process with the solution expressed as a weighted sum of the same set of basis vectors.
After integrating by parts, the $i^{th}$ equation is
\begin{equation}
\label{eq:land3}
\int \limits_{\Omega} d{\vec{x}} \; \phi_i \; \frac{df\left(\vec{x} \right)}{dt} = - \int \limits_{\Omega} d{\vec{x}} \; \nabla \phi_i \cdot \left( \nabla \cdot \mathbf{D}\left(\vec{x} \right) \nabla - \nabla \cdot \mathbf{K}\left(\vec{x} \right) \right) f\left(\vec{x} \right).
\end{equation}

The exact Jacobian of the system is dense, however an approximation that ignores linearizing the inner integrals is sparse.
That is, at each point $\vec{x}$ one integrates over all domains to compute $\mathbf{D}\left(\vec{x} \right)$ and $\mathbf{K}\left(\vec{x} \right)$ and then discretizes $\nabla \cdot \mathbf{D}\left(\vec{x} \right)\nabla $ and $\nabla \cdot \mathbf{K}\left(\vec{x} \right)$.
%Note, after $A_h\left[u_h\right]$ is computed a matrix-vector produce $A_h\left[u_h\right]u_h$ is correct and only correct for the one $u_h$.

\subsection{Data and work partitioning model}
\label{sec:data}

In matrix form, with $S$ species and letting $u = \left[ f_1, f_2, ... , f_S \right]^T$, the system can be written as an $S \times S$ block matrix, $A$, defined by $A\left[u\right]\left(\alpha,\beta\right) \equiv C\left[f_\alpha,f_\beta\right]_{\alpha\beta}$ from (\ref{eq:landau1}).
An implicit time integrator, backward Euler for example, evolves a state $u^{k}$ at time $t$ to the state $u^{k+1}$ at time $t + \Delta t$ according to $u^{k+1} \xleftarrow{} u^{k} + \Delta t A\left[u^{k+1}\right] u^{k+1}$, which requires a nonlinear solve of $\left(I - \Delta t A\left[u^{k+1}\right] \right) u^{k+1} = u^{k}$ for $u^{k+1}$.
The finite element discretization results in the nonlinear algebraic system $\left (M_h + \Delta t A_h\left[u_h\right]\right)u^{k+1}_h = u^k_h$, with the finite element {\textit{mass}} matrix $M_h$ (see \S III in \cite{Hirvijoki2016} for details).
The mass matrix is added in a separate kernel due to the PETSc time stepper interface design. % and that the Jacobian matrix is computed in a residual calculation when the state changes. -- OPTIONAL

A Newton iteration first computes the residual with the current solution $u$, $r \xleftarrow{}u^k - A_h\left[u\right]u$, then solves for the correction $\delta$ in $\left (M + \Delta t A_h\left[u\right]\right)\delta = r$ and updates the solution $u \xleftarrow{} u + \delta$.
\begin{figure}[h!]
    \centering
    \includegraphics[width=1\linewidth]{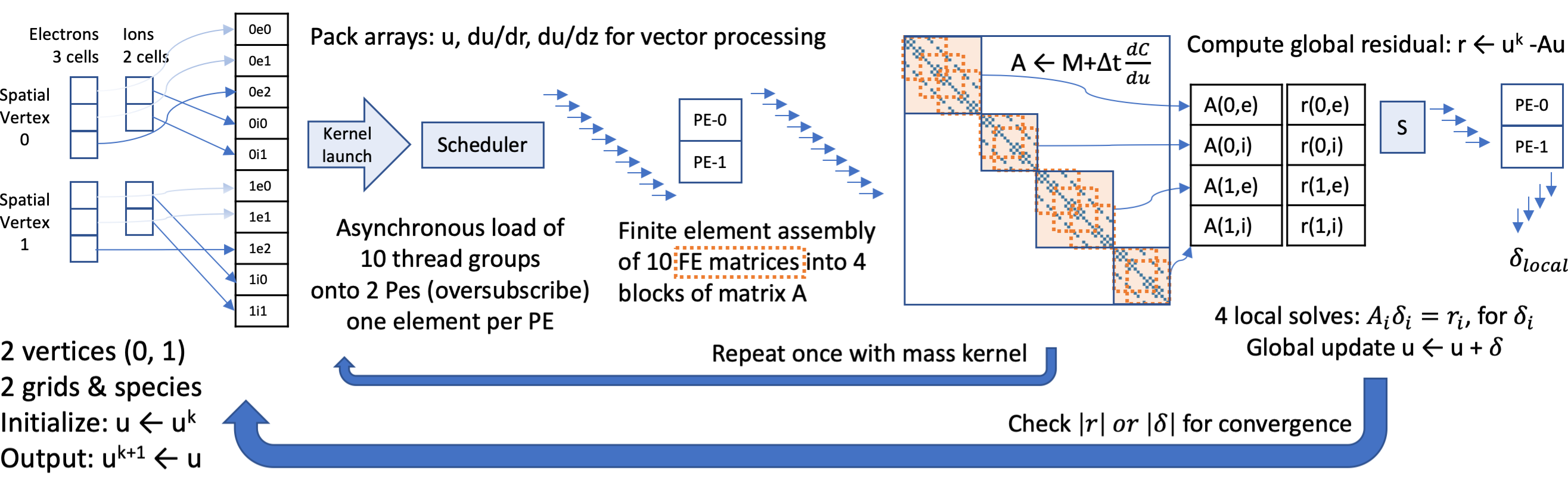}
    \caption{Diagram of one time step on a generic accelerator}
    \label{fig:arch}
\end{figure}
Figure \ref{fig:arch} diagrams a backward Euler time step with the nonlinear solver, batched matrix construction and linear solver with two spatial vertices, a two species plasma, a three cell electrons (e) and two cell ions (i) mesh on a generic accelerator with asynchronous scheduler (S).
All work and data remain on the device throughout the full collision time advance.

\subsubsection{Jacobian and mass matrix construction}
\label{ssec:landau_st}

The algorithm for computing the Jacobian matrix of (\ref{eq:landau1}) can be expressed as six nested loops (Algorithm 1, \cite{AdamsHirvijokiKnepleyBrownIsaacMills2017}), and implemented with a naive kernel (Algorithm 3, \cite{AdamsHirvijokiKnepleyBrownIsaacMills2017}) and optimized by computing sub-expressions and loop hoisting (Algorithm 1, \cite{Adams2022a}).
The data structures are optimized for vector processing (Algorithm 2, \cite{AdamsHirvijokiKnepleyBrownIsaacMills2017}).
Our strategy for partitioning data and work is to compute each element's matrix with one Kokkos thread group, or PE, and assemble the global Jacobian matrix on the device with the COO interface in PETSc \cite{mills2024petsctao}, as illustrated in Figure \ref{fig:arch}.

The finite element mass matrix is linear and does not change during the simulation because the basis functions on different grids are trivially orthogonal. % separate grids
One can compute the mass matrix once, on the CPU for convenience, copy it to the GPU and add it to the shifted Jacobian matrix.
We found that while this approach was a bit faster with small number of batches the memory bandwidth demands with large batch sizes resulted in poor performance and recomputing the mass term on-the-fly on the GPU was faster and used less memory.
%Thus, the matrix for the linear solver consists of the (shifted) Jacobian plus the mass matrix as diagrammed in Figure \ref{fig:arch}.
%The performance characteristics of each of these two parts is very different and they are analyzed separately. 

\subsubsection{Shared memory}
The batch solvers offer the opportunity to place data into shared memory explicitly instead of relying on the cache system to pull data from global memory.
Shared memory tends to be partitioned into a user managed partition and a system cache partition.
PETSc, like the Kokkos Kernels batch solvers, fills the user shared memory space with as many Krylov work vectors as will fit, prioritizing vectors appropriately, and relying on the system to manage the rest.
We observe that the cache system is effective, with only a small performance increase from explicit use of shared memory.
The matrix construction also explicitly places data in shared memory, such as the accumulation of integrals for $\mathbf{D}$ and $\mathbf{K}$ in (\ref{eq:land2}), as is described in \S III.D and \S III.E of \cite{Adams2022a} .

\subsection{Coordinate systems}

The Landau integral is inherently three dimensional, but in a strong magnetic guide field, a gyrokinetic approximation allows for the use of cylindrical coordinates, $\vec{x} = \left ( r, z \right)$, to reduce the $3D$ problem to a $2D$ computation, or in velocity space $2V$, and (\ref{eq:land2}) is modified accordingly (\S III.A \cite{Adams2022a}).
The $z$ coordinate is aligned with the magnetic field and is referred to as $v_\parallel$.
Likewise $r$ is referred to as $v_\perp$. 
Both the $2V$ and full $3V$ models are investigated in this report.
The $3V$ model is required for extension to relativistic regimes \cite{Beliaev_Budker_1956,Braams-Karney:PRL1987,PhysRevE.99.053309}, but $2V$ is the focus of performance optimization and is currently practical \cite{Hager2016}.

\subsection{The Vlasov-Maxwell-Landau system}

%To provide context and to motivate batching, t
While the collision operator in this paper can be used as-is, it is intended for use within a fully kinetic model, such as the \textit{Vlasov-Maxwell-Landau} system, the fundamental model of magnetized plasmas \cite{Vlasov1968,landau1936kinetic}.
The density for each species is evolved in phase space, using $\vec{x}$ for the spatial coordinate and $\vec{v}$ for the velocity coordinate, with up to three \textit{configuration} space dimensions and three velocity space dimensions (6D or $3X + 3V$) according to
\begin{equation*}
\frac{df_\alpha}{dt} \equiv
\frac{\partial f_\alpha}{\partial t} + \frac{\partial  \vec{x}}{\partial t}
\cdot \nabla_x f_\alpha + \frac{\partial \vec{v}}{\partial t} \cdot \nabla_v f_\alpha = \sum_{\beta} C\left[f_\alpha,f_\beta\right]_{\alpha\beta}.
\end{equation*}
Maxwell's equations provide electro-magnetic acceleration forces in the $\partial \vec{v}/\partial t$ term.

The evolution of the system is split between a global symplectic time evolution of the Vlasov-Maxwell system and a velocity space evolution of the collision operator. 
This paper focuses on the performance and verification of the collision operator.
Our test problems, which are examples in PETSc, assume $\nabla_x = 0$ and $\partial \vec{v}/\partial t = a = 0$.
However $\partial \vec{v}/\partial t$ terms are folded into the collision operator for a plasma resistivity verification test where a constant electric field is applied, inducing a current that is measured and the effective resistivity compared to the NRL Plasma Formulary \cite{Adams2022a}.

\section{Multiple grids and batching}
\label{sec:batching_mg}

Separate grids for each species simplify meshing because only a single scale needs to be resolved for the near-Maxwellian distributions common in plasmas, and adaptive meshing is less critical than if a single grid is used \cite{AdamsHirvijokiKnepleyBrownIsaacMills2017}.
Additionally, species with similar thermal speeds, such as with many ionization states of impurities, can share a grid and amortize the cost of the Landau tensor \cite{Hirvijoki2016}, which is substantial in $2V$ (see Appendix in \cite{Hirvijoki2016}).

Figure \ref{fig:3grids} shows three block-structured AMR grids used for the 10 species example in \S\ref{sec:throughput_perf}, with an electron grid, a light ion and heavy ion grid, plotted with a Maxwellian distributions in the axisymmetric coordinate system.
Note the different scales on each of the grids, otherwise the images are identical.

\begin{figure}[h!]
\begin{center}
\begin{subfigure}{0.25\linewidth} %\centering
    \includegraphics[width=\linewidth]{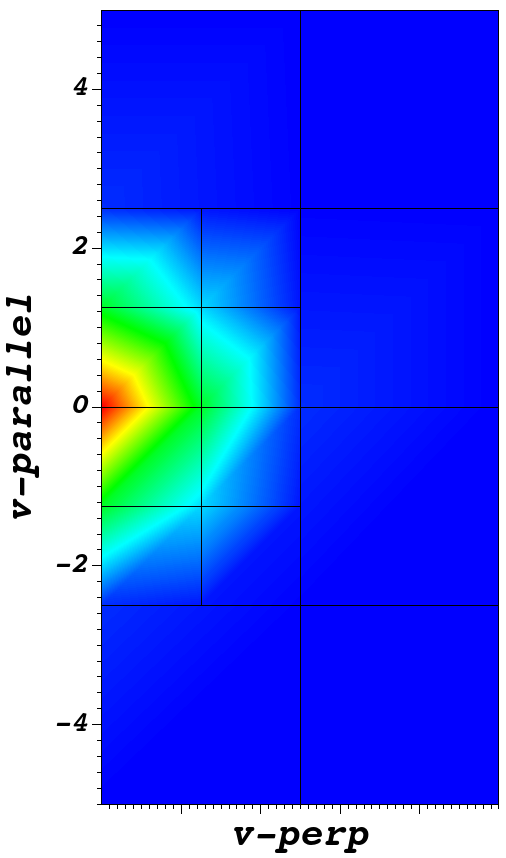}
    \caption{electrons}
    \label{fig:amr-e}
\end{subfigure}
\begin{subfigure}{0.285\linewidth} %\centering
    \includegraphics[width=\linewidth]{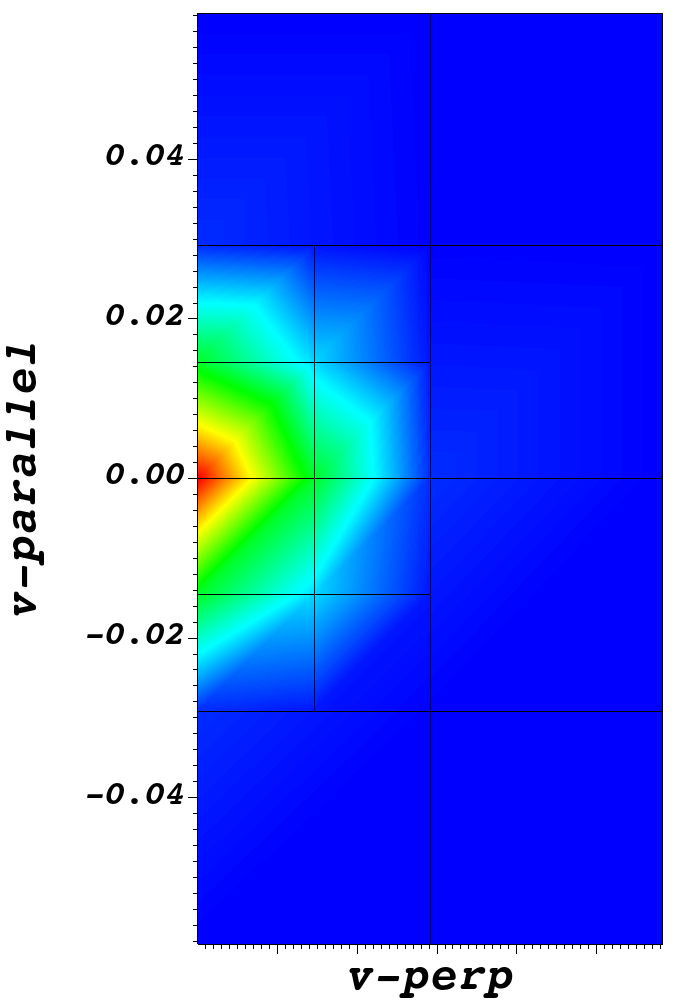}
    \caption{Deutarium}
\end{subfigure}
\begin{subfigure}{0.25\linewidth} %\centering
    \includegraphics[width=\linewidth]{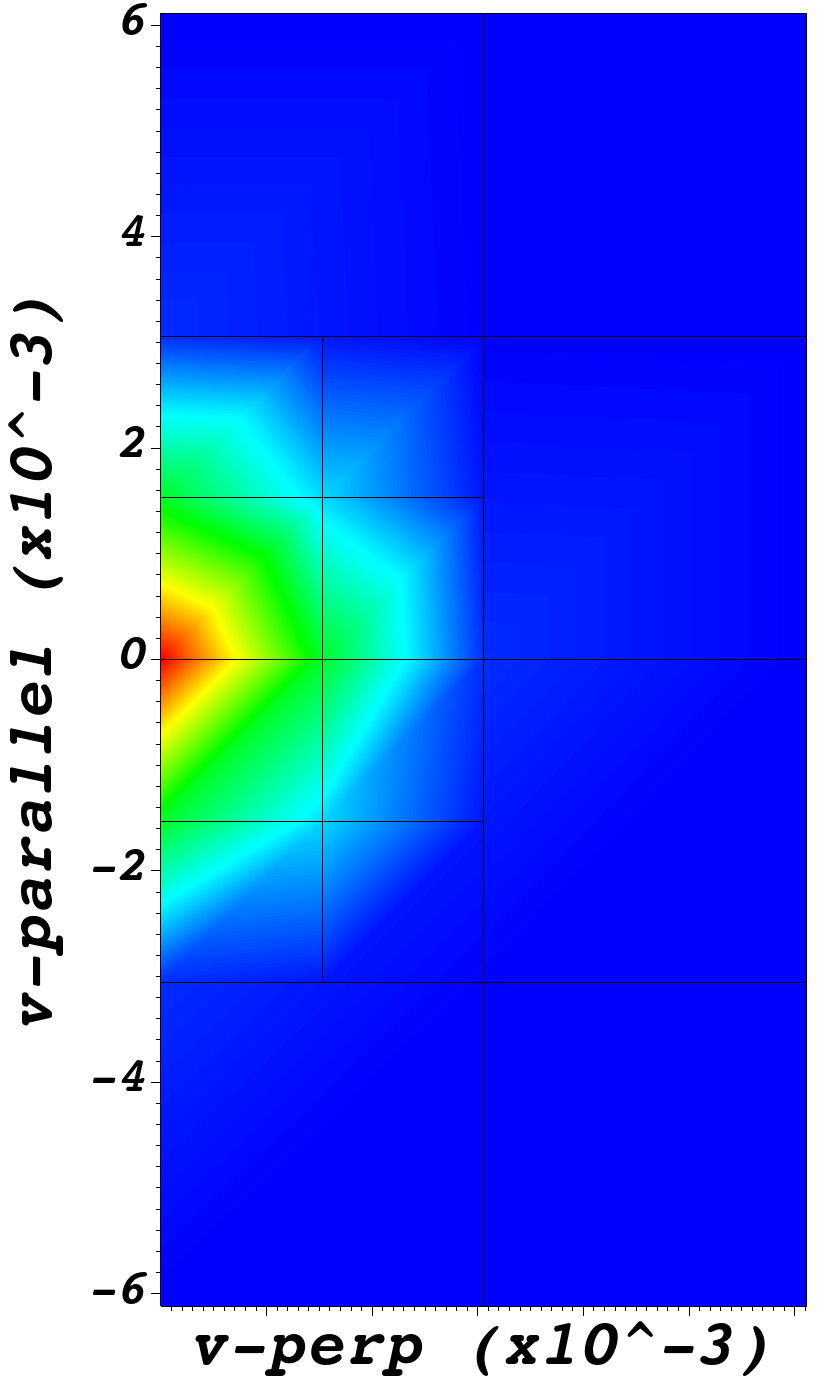}
    \caption{Tungsten}
\end{subfigure}\vspace{-0.15cm}
\caption{Three grid example with Maxwellian distributions with scales for each species group. (Linear interpolation in Visit results in visualization artifacts)}
\label{fig:3grids}
\end{center}
\end{figure}
\vspace{-0.35cm}

\subsection{Batching and aggregation}
\label{sec:batching}

Kinetic applications commonly use operator split time integrators, where the symplectic Vlasov system evolves the full phase space distribution and the metric collision operator evolves the velocity space distribution separately at each spatial grid point.
%Tensor product grids, with a $3X$ spatial grid of the application domain and a $2-3V$ velocity space grid at each spatial grid ``point", are commonly used \cite{Hager2016}.
%Thousands or more collision advances run -- independently -- on each device in each collision step.
This algorithm has a high degree of parallelism and is well-suited to modern hardware with large numbers of simple processing elements (PEs). % where, by definition, a fast synchronization primitive is available. 
\textit{Batching} refers to writing algorithms in a single, or a few, kernels that run asynchronously on PEs so as to fully exploit available parallelism in the algorithm.
%Batching is similar to \textit{loop fusion}.
In the Landau operator the computation and assembly of element ``stiffness" matrices and the subsequent linear solve are amenable to batching.

An alternative to batching is to \textit{aggregate} the many small systems into a single system that can use sparse linear algebra operators in vendor libraries or higher level libraries like PETSc.
Aggregation requires block diagonal matrices as illustrated in Figure \ref{fig:arch}.
The aggregated methods use separate kernel launches for each primitive operation such as a matrix-vector product. %, but are still run on the device.
The time integrators use high-level operations that are not compute intensive, are not iterative, and with adaptive time stepping require global synchronization, and therefore aggregation is appropriate.
Nonlinear solvers also use high-level operations and can be aggregated effectively, but the asynchronous processing of batching would benefit problems 
 with nonhomogeneous vertex problems.
We currently use PETSc's (aggregated) nonlinear solvers -- batching is the subject of future work.

\subsection{Linear solvers and batching}
\label{ssec:batching_solvers}

%The Jacobian matrix is block diagonal, with a block for each species in addition to vertex blocks, which adds another potential dimension to batching.
Iterative solvers lend themselves to both aggregated and batched solvers.
The sparse matrix-vector product kernels in iterative solvers are amenable to vector processing, and with few data dependencies they are well-suited to modern accelerators.
Aggregation solvers compute residuals, solve linear systems, and update solutions with traditional high-level sparse linear algebra on systems composed of aggregated small systems.
Batch solvers require writing the entire solver and preconditioner in device kernel code.
Batch solvers have several advantages in both performance and accuracy: the entire solve requires a single kernel launch, each solve can run asynchronously, exiting the kernel as soon as it converges to be immediately replaced with another solve by the scheduler, and (unwanted) global synchronization is avoided.
Aggregate solvers merge the systems together with a single (L2 norm) convergence test, which obscures the residual of each system.
Batch solvers maintain the semantics of the Krylov method for each system.
Batched sparse solvers are available in the PETSc, Kokkos Kernels and Ginkgo libraries \cite{Liegeois2023,Kashi_dhruva}.

In addition to code reuse, the aggregation approach has the advantage of being able to run effectively on a larger range of problem sizes because an individual linear system can run on just part of a PE or multiple PEs, transparently.
Note, Kokkos Kernels and Ginkgo batch solvers support scaling down with multiple problems per thread group, but the PETSc batch solver does not.
For scaling up, NVIDIA provides \textit{cooperative groups} since CUDA-9 and the new \textit{Hopper} architecture introduces hardware support for inter-PE communication called a \textit{thread block cluster} that should allow for Kokkos to increase the size of its thread groups to use 16 SMs in the future.
The $3V$ solves in \S\ref{sec:throughput_solver} can be seen to suffer from this scaling up problem.

\section{Landau time advance performance}
\label{sec:throughput_perf}

The vast majority of the time in the full Landau time advance is spent in matrix construction and linear solves (e.g., Table \ref{tab:parts-Perlmutter-2V}), and the high-level timings in this section, and \S\ref{sec:perfanisotropic}, focuses on these two phases.
\S\ref{sec:utilization} analyzes low-level hardware utilization.
This section uses a 10 species model problem to investigate throughput with respect to batch size and $2V$ vs $3V$ in \S\ref{sec:throughput_model}, and component times with batched and aggregate linear solves in \S\ref{sec:componets_throughput}. 
\S\ref{sec:perfanisotropic} investigates component times with a two species anisotropic relaxation test with respect to order of the finite space with tensor and simplex elements.
%All performance test on conducted with a single NVIDIA A100 and AMD MI250X nodes, and a test harness replicates the problem on each MPI process with a give \textit{batch size}.
The test harness replicates the model problem to create a batch of problems in each MPI process, to mimic an application.
Each NVIDIA GPU or AMD GCD on a compute node is driven by one MPI process.

%The performance tests consider, first and foremost, batch size
%
%One could use the throughput of time steps or nonlinear solves, but the cost of a time step and the number of nonlinear solves depends on the time integrator and user parameters such as the time steps size, time accuracy, and the nonlinear solver tolerance, which determines the degree of energy conservation.
%For a given problem, the total number of Newton iterations is somewhat invariant to time step size because as the time step is increased, the number of iterations in the nonlinear solver increases because there is more nonlinearity and the mollifying effect of the mass term is reduced. 

\paragraph{A 10 species model problem}

The model problem is a deuterium plasma with the addition of eight species of tungsten with different ionization states.
This resembles a fusion energy science benchmark for plasma models with impurities.
The electrons and deuterium each have a dedicated grid and the tungsten species share one grid, for a total of three block-structured AMR grids as shown in Figure \ref{fig:3grids}.
One level of AMR refinement about the origin is used with a $4 \times 2$ and a $4 \times 4 \times 4$ initial grid, in $2V$ and $3V$ respectively, resulting in 14 (Q3 elements) in $2V$ and 120 (Q2 elements) in $3V$.
We have observed that these grids are sufficient to converge a plasma resistivity test to within about $1\%$ of the fully converged state \cite{Adams2022a}.
Each problem has 10 linear systems, one for each species, with 142 equations and $22.2$ average non-zeros per row, in $2V$, and 1,045 equations with an average of $54.0$ non-zeros per row in $3V$.
Each distribution is initialized with a Maxwellian with a given thermal temperature.
An implicit backward Euler time integrator with electron temperature initialized to twice that of the ions is run toward equilibrium for ten time steps.

\subsection{Throughput experiment}
\label{sec:throughput_model}

This section analyzes the throughput of the entire Landau time advance as well as that of the linear solvers, with respect to batch size for both the $2V$ and full $3V$, and for batched and aggregate solvers.
Newton throughput is defined as the number of Newton iterations in the entire simulation times the batch size per device times the number of devices per node, divided by the \textit{total run time}, which includes the Jacobian and mass matrix construction, the linear solve, and some linear algebra in the control logic of the time integrator and nonlinear solver, with a (CPU) setup phase that is amortized by the solve phase.
Given that all tests use the same Landau solver with a slightly different linear solver, one could use many quantities of interest, such as flop rate, but Newton iterations is chosen because it is relevant to application scientists.
Solver throughput is defined similarly, with the number of linear solves (with $10-15$ per nonlinear solve) divided by the time spent in the linear solve phase only.

%The density in each item in the batch is increased slightly to add some nonuniformity to better mimic an application. 
Each of these problems requires a linear solve per species, resulting in ten batched linear solves for each problem in the batch.
%To mimic variability in a real application, the density and hence collision frequency of each successive problem in a batch are varied within a range of about $10\%$.
Two linear solvers are considered: a batched TFQMR solver in PETSc, written in Kokkos, and an aggregated TFQMR solver that uses Kokkos Kernels linear algebra primitives within the PETSc framework.
Jacobi preconditioning is used throughout.
Two architectures are examined:
\begin{itemize}
    \item one node with four NVIDIA A100 Tensor Core GPUs with 256GB of memory, the \textit{Perlmutter} machine at NERSC and, 
    \item  one node with four AMD MI250Xs, each with 2 Graphics Compute Dies (GCDs), with ROCm 5.1, the \textit{Crusher} machine at ORNL.
\end{itemize}

\subsubsection{Newton iteration throughput}
\label{sec:throughput_newt}

Table \ref{tab:Newton-throughput-2V} shows the $2V$ Newton iteration throughput, and Table \ref{tab:Newton-throughput-3V} shows the $3V$ data, as a function of batch size with the batch linear solver and aggregated linear solver for the A100 and MI250X.

\begin{table}[h!]
   \begin{center}
     \begin{tabular}{ c  c  }
       \includegraphics[width=0.46\linewidth]{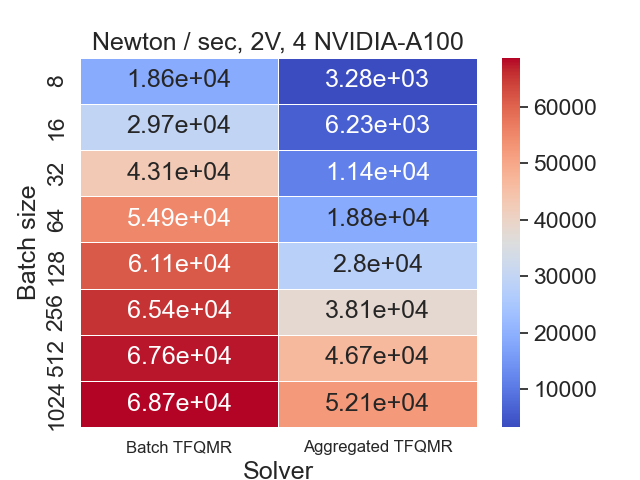}
        & 
       \includegraphics[width=0.46\linewidth]{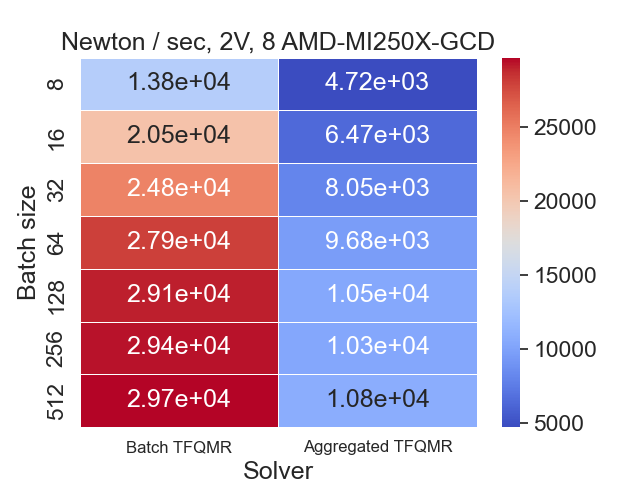}
     \end{tabular} 
     \vspace{-0.15cm}
   \caption{2V Newton throughput: NVIDIA A100 (left) and AMD MI250X (right)} 
   \label{tab:Newton-throughput-2V}
   \end{center}
\end{table}
%\vspace{-1.cm}
\begin{table}[h!]
   \begin{center}
     \begin{tabular}{ c  c  }
       \includegraphics[width=0.46\linewidth]{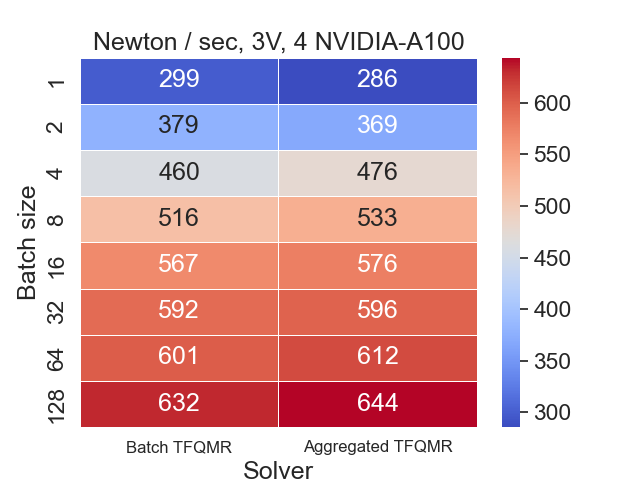}
        & 
       \includegraphics[width=0.46\linewidth]{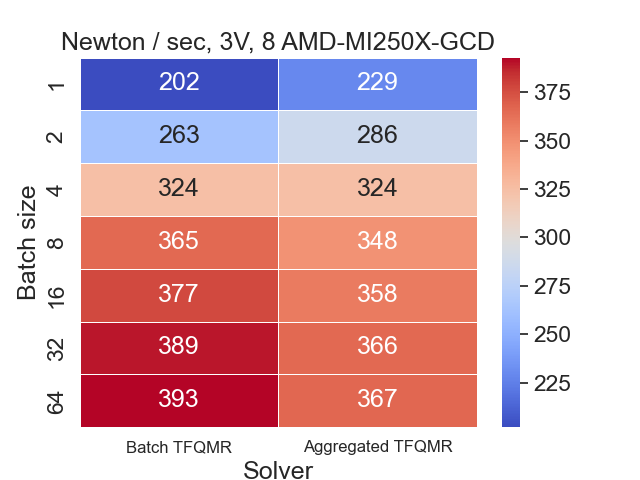}
     \end{tabular}
     \vspace{-0.15cm}
   \caption{3V Newton throughput: NVIDIA A100 (left) and AMD MI250X (right)}
   \label{tab:Newton-throughput-3V}
   \end{center}
\end{table}

The next section investigates the linear solver throughput, but this data shows that the A100 is more than $2x$ faster than the MI250X in $2V$ and about $50\%$ faster in $3V.$
Note that the GPU is pretty well saturated with a batch size of 128 in $2V$ and 16 in $3V$, and this corresponds to 57,344 and 42,240 integration points per GPU, and thus $2V$ and $3V$ saturate with a similar number of total integration points.

\subsubsection{Linear solver throughput}
\label{sec:throughput_solver}

Table \ref{tab:solver-throughput-2V} shows the $2V$ linear solver throughput, and Table \ref{tab:solver-throughput-3V} shows the $3V$ throughput data, as a function of batch size with the batch linear solver and aggregated linear solver for the A100 (left) and MI250X (right).
\begin{table}[h!]
   \begin{center}
     \begin{tabular}{ c  c  }
       \includegraphics[width=0.46\linewidth]{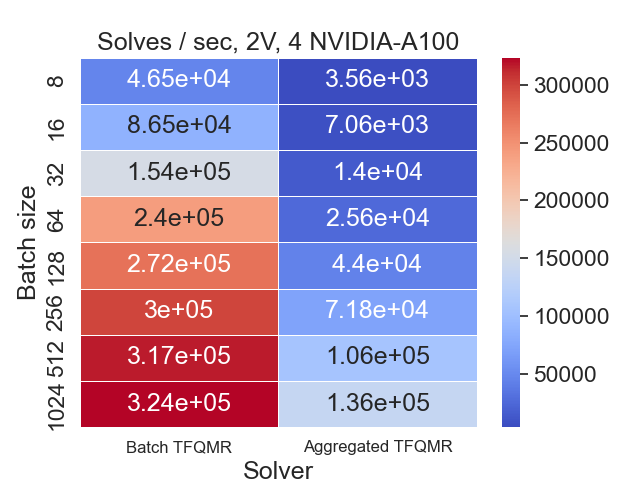}
        & 
       \includegraphics[width=0.46\linewidth]{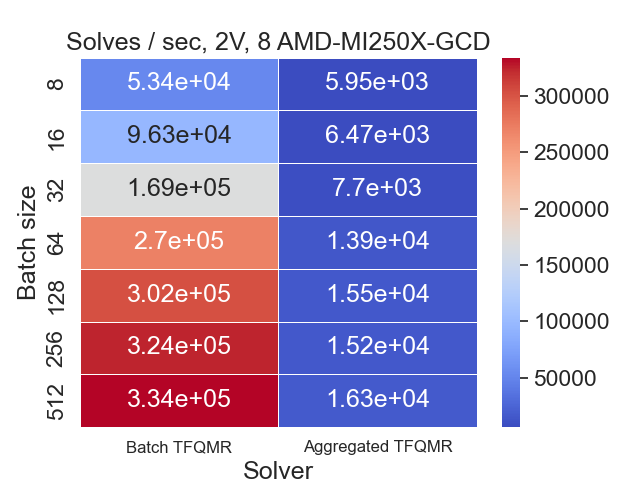}
     \end{tabular}%\vspace{-0.15cm}
   \caption{2V Linear solver throughput: NVIDIA A100 (left) and AMD MI250X (right)}
   \label{tab:solver-throughput-2V}
   \end{center}
\end{table}
%\vspace{-1.cm}
\begin{table}[h!]
   \begin{center}
     \begin{tabular}{ c  c  }
       \includegraphics[width=0.46\linewidth]{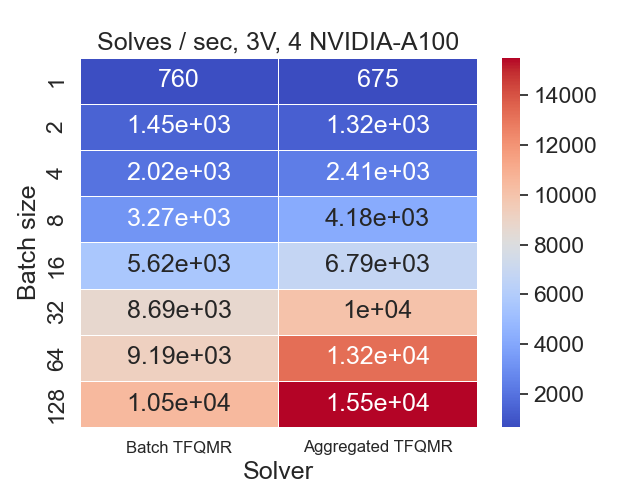}
        & 
       \includegraphics[width=0.46\linewidth]{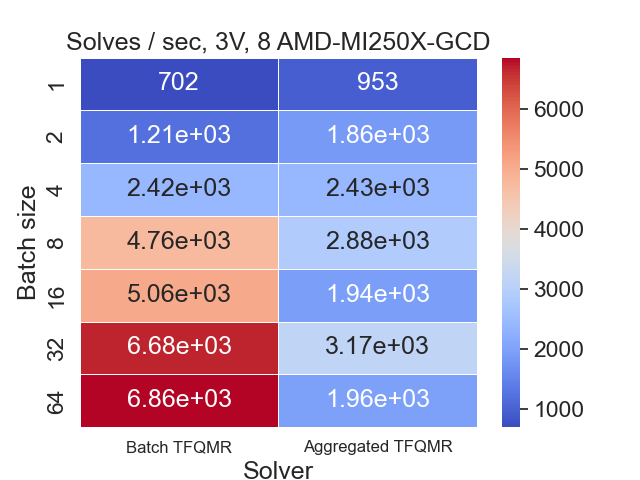}
     \end{tabular}%\vspace{-0.15cm}
   \caption{3V Linear solver throughput: NVIDIA A100 (left) and AMD MI250X (right)}
   \label{tab:solver-throughput-3V}
   \end{center}
\end{table}
 
This data shows that the batched TFQMR solver over $20x$ faster than the aggregated solver in $2V$ and the MI250X and over $2x$ faster on the A100, and that the two batch solvers have similar performance.
The A100 batch solver is almost $2x$ faster in $3V$, but the aggregated solver is faster than the batch solver on the A100. 
The batch solver is $3.5x$ faster than the aggregate solver on the MI250X.
This under-performance of the batched solver in $3V$ illustrates the difficulty in fitting the problem to a fixed hardware resources as discussed in \S\ref{ssec:batching_solvers}.
The linear solver also does not saturate as clearly as the Jacobian dominated Newton iteration throughput.

\subsection{Component times}
\label{sec:componets_throughput}

Tables \ref{tab:parts-Perlmutter-2V} and \ref{tab:parts-Perlmutter-3V} show the component times on the A100 in $2V$ and $3V$, respectively, including mass matrix  (``Mass"), Landau Jacobian (``Jacobian"), linear solver (``Solve"), the total time and the sum of the Krylov iterations for the aggregate solver and the sum of the maximum number of iterations of any system for the batch solver (see \S\ref{sec:batch_its} for the batch solver iteration count distribution).
Tables \ref{tab:parts-Crusher-2V} and \ref{tab:parts-Crusher-3V} show the same data on the MI250X in $2V$ and $3V$, respectively.
\begin{table}[h!]
\centering
\caption{2V Component times (batch size = 256), NVIDIA-A100}
\label{tab:parts-Perlmutter-2V}
\begin{tabularx}{\textwidth}{lrrrrl}
\toprule
Component &  Jacobian &  Mass &  Solve &  Total & Krylov iterations \\
\midrule
Batch TFQMR    &      1.57 &  0.22 &   0.58 &   2.44 &      3,648 \\
Aggregated TFQMR &      1.57 &  0.22 &   1.76 &   3.69 &      4,015 \\
\bottomrule
\end{tabularx}
\end{table}
\begin{table}[h!]
\centering
\caption{3V Component times (batch size = 32), NVIDIA-A100}
\label{tab:parts-Perlmutter-3V}
\begin{tabular}{lrrrrl}
\toprule
Component &  Jacobian &  Mass &  Solve &  Total & Krylov iterations \\
\midrule
Batch TFQMR    &     29.69 &  3.00 &   2.33 &  35.08 &      2,785 \\
Aggregated TFQMR &     29.67 &  3.00 &   1.51 &  34.31 &      2,326 \\
\bottomrule
\end{tabular}
\end{table}
\begin{table}[h!]
\centering
\caption{2V Component times (batch size = 128), AMD-MI250X-GCD}
\label{tab:parts-Crusher-2V}
\begin{tabular}{lrrrrl}
\toprule
Component &  Jacobian &  Mass &  Solve &  Total & Krylov its \\
\midrule
Batch TFQMR    &      4.28 &  0.29 &   0.49 &   5.08 &      3,642 \\
Aggregated TFQMR &      4.31 &  0.29 &   9.54 &  14.10 &      4,011 \\
\bottomrule
\end{tabular}
\end{table}
\begin{table}[h!]
\centering
\caption{3V Component times (batch size = 64), AMD-MI250X}
\label{tab:parts-Crusher-3V}
\begin{tabular}{lrrrrl}
\toprule
Component &  Jacobian &  Mass &  Solve &  Total & Krylov iterations \\
\midrule
Batch TFQMR    &    168.16 & 18.07 &  11.28 & 196.81 &      2,796 \\
Aggregated TFQMR &    168.62 & 18.06 &  39.51 & 210.84 &      2,326 \\
\bottomrule
\end{tabular}
\end{table}

%The linear solves are subdominant in $2V$ and insignificant in $3V$, which is expected because the $\Order{N^2}$ work complexity of the Jacobian algorithm problem size increases faster than that of the linear solver.
The balance of time between the Jacobian and linear solver is skewed toward the Jacobian construction on the AMD relative to the NVIDIA data.
This is partially due to the AMD accelerator outperforming the NVIDIA node in the linear solver, but the MI250X node underperforms in the matrix construction.
%It should be noted that at the time of these experiments that the MI250X and ROCm was a less mature product than the A100.
%This is likely at least partially a function of Kokkos being more highly optimized for NVIDIA hardware, but this is a subject of future work.
%Note, the mass matrix creation is essentially only finite element assembly and sparse matrix assembly and thus the Jacobian time minus the mass time indicates the cost of the Landau kernel.

\subsubsection{Batch solver iteration distributions}
\label{sec:batch_its}

The Krylov iteration counts for the batched solvers in Tables \ref{tab:parts-Perlmutter-2V}, \ref{tab:parts-Perlmutter-3V}, \ref{tab:parts-Crusher-2V} and \ref{tab:parts-Crusher-3V} obscures the distribution of iterations within the 10 linear species solve.
Table \ref{tab:ave-nits} shows average iterations per linear solve for aggregate (Agg.) and batch (Bat.) solvers in $2V$ and $3V$, with batch solver iteration count averages for each species 1-10, with one time step of the test problem. 
The $2V$ batch size is 256 and the batch size in $3V$ is 32.
\begin{table}[h!]
\centering
\caption{Average linear solver iteration counts: aggreate, batched and each of ten species in the batched solve}
\label{tab:ave-nits}
\resizebox{\columnwidth}{!}{%
\begin{tabular}{lrrrrrrrrrrrr}
\toprule
V &  Agg. &  Bat. &    1 &    2 &    3 &    4 &    5 &    6 &    7 &    8 &    9 &   10 \\
\midrule
2 &  23.9 &  11.8 & 23.6 & 10.0 & 11.2 & 11.2 & 10.1 & 10.1 &  9.9 & 11.0 & 10.0 &  9.2 \\
3 &  16.6 &  14.8 & 14.0 & 15.9 & 17.0 & 16.2 & 15.8 & 15.3 & 15.4 & 15.5 & 15.1 & 14.8 \\
\bottomrule
\end{tabular}
}
\end{table}

In $2V$ the electrons (first species) have a much higher iteration count, likely due to the higher speed of electrons (see Figure \ref{fig:3grids}) and other implicit Landau solvers have noticed this phenomenon \cite{Kashi_dhruva}.
The $3V$ data, however, does not exhibit this behavior, which is not understood.

\section{Unstructured mesh generation with simplex and tensor elements}
\label{sec:meshing}

The finite element formulation and implementation in this work supports unstructured simplex (triangles) and tensor cell (quadrilaterals and hexahedra) meshes as well as the block-structured AMR meshes used in \S\ref{sec:throughput_perf}. 
%The nodal positions of the block structured AMR meshes shown in Fig. \ref{fig:3grids} have been hand optimized to provide high accuracy with minimal degrees of freedom. 
This section focuses on simplex mesh generation and presents the unstructured quadrilateral meshes newly available.
Simplex meshes are read from a file and quadrilateral mesh can be similarly input, but the solver provide simple quadrilateral mesh generation (\S\ref{ssec:quads}).

\subsection{Simplex mesh generation}
\label{sec:meshing_simp}
%With the unstructured meshes, geometric partitioning is required to optimize accuracy. 
The simplex meshes are constructed on a semicircle centered at $(0,0)$ with a radius of one. They are scaled within the solver to obtain the final dimensions. Unstructured meshes have been generated using simmetrix-simmodsuite version 18.0-220913 \cite{simodsuite}. 

\begin{table}[h!]
\begin{center}
\caption{Mesh quality metrics for each of the unstructured meshes.}
\label{tab:mesh-statistics}
\begin{tabular}{cc||rr|rr}
\toprule
 &  &\multicolumn{2}{c|}{Aspect Ratio} & \multicolumn{2}{c}{Skew} \\
\# cells & P & Max  & Median & Max & Median\\
\midrule
144 & 2 & 2.618 & 1.512 & 0.552 & 0.161 \\
68 & 3 & 2.926 & 1.762 & 0.650 & 0.246 \\
24 & 4 & 2.801 & 1.304 & 0.254 & 0.097 \\
\bottomrule
\end{tabular}
\end{center}
\end{table}

\begin{figure}[h!]
\begin{center}
\begin{subfigure}{0.32\linewidth} \centering
    \includegraphics[width=\linewidth]{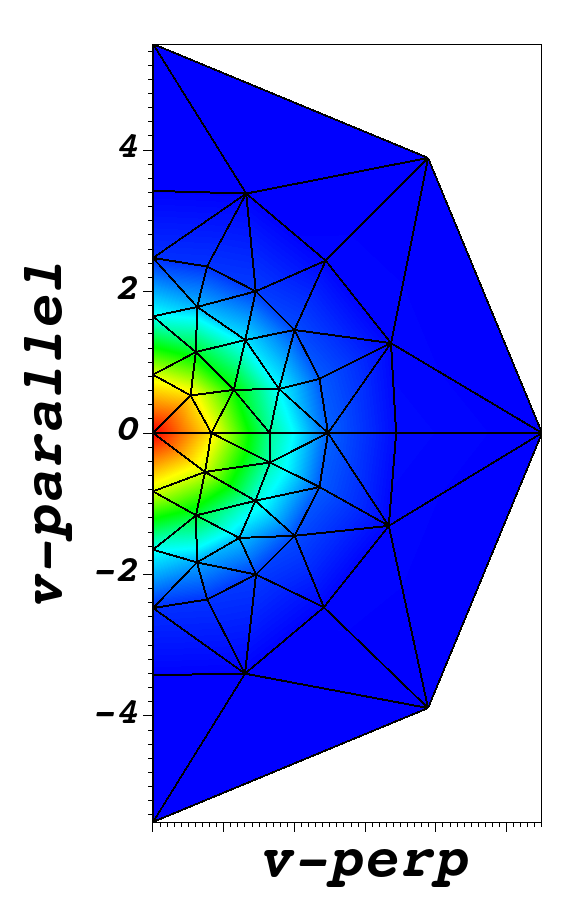}
    \caption{$P2$ grid}
    \label{fig:circle-p2}
\end{subfigure}
\begin{subfigure}{0.31\linewidth} \centering
    \includegraphics[width=\linewidth]{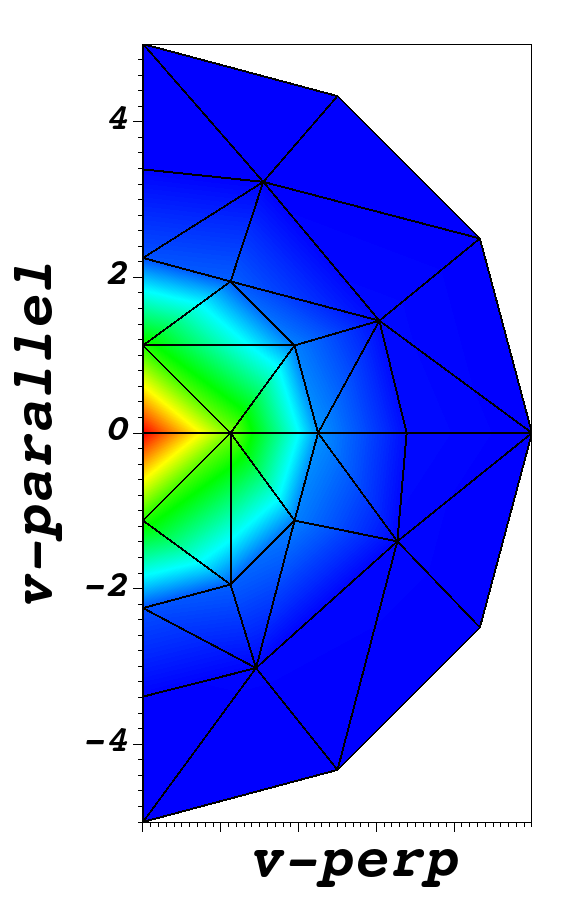}
    \caption{$P3$ grid}
\end{subfigure}
\begin{subfigure}{0.32\linewidth} \centering
    \includegraphics[width=\linewidth]{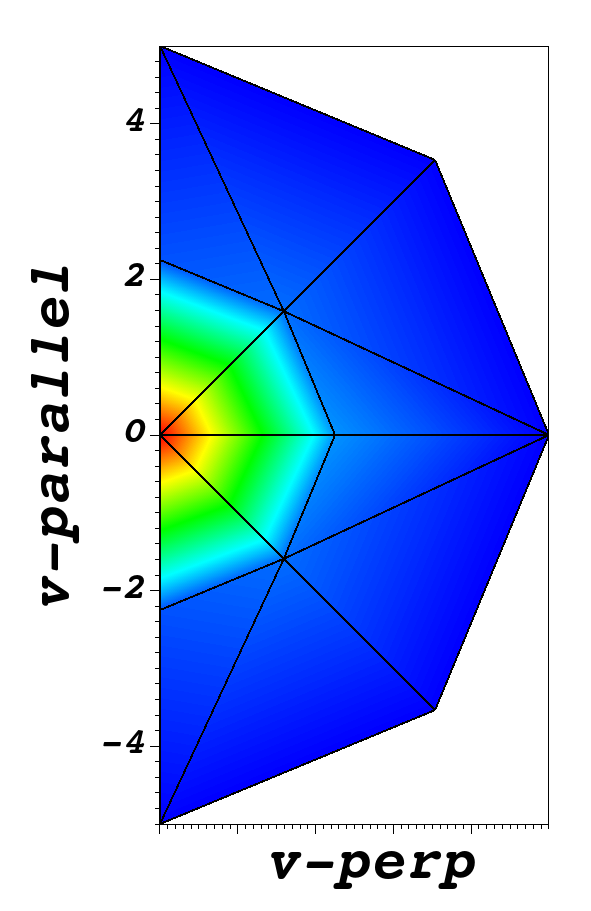}
    \caption{$P4$ grid}
\end{subfigure}
\caption{Semi-circular grids for the three orders of finite elements ($P2$, $P3$ and $P4$) tested with Maxwellian distribution of electrons in units of the thermal speed (Linear interpolation in Visit results in visualization artifacts)}
\label{fig:circle}
\end{center}
\end{figure}

Each model is partitioned at $z=0$ which promotes a more symmetric mesh construction. Additionally, the model is partitioned at $r=0.45$. We found that high aspect ratio elements near the boundary of the Maxwellian distribution caused a loss in accuracy. Mesh construction required a balance between the number of elements in the outer region ($r>0.45$), largely controlled by gradation rate, and aspect ratio of these elements. We found that prioritizing the aspect ratio criteria over other element shape metrics during the mesh optimization and smoothing phases improved the overall mesh quality and solution accuracy. The shape quality metrics for the meshes used herein (Fig. \ref{fig:circle}) are given in table \ref{tab:mesh-statistics}, where the aspect ratio is the ratio of the length of the longest triangle edge to the shortest triangle edge and the skew is defined as
\begin{equation}
    \text{Skew} = \frac{A_\text{opt} - A}{A_\text{opt} }.
    \label{eq:area-skew}
\end{equation}
For 2D simplices the optimal area \(A_\text{opt} = 3\sqrt{3}/4 R^2\) where \(R\) is the circumcircle radius corresponds to the equilateral triangle. The skew ranges from \(0\) (equilateral) -- \(1\) (degenerate), where quality meshes of complex geometries typically have values of \(\text{Skew} < 0.8\).

These meshes are read into the solver from a file. % as opposed to the internally generated block-structured AMR meshes shown in Figure \ref{fig:3grids}.
The three meshes used for the anisotropic relaxation model (\S\ref{sec:verify}) with the initial electron distribution is shown in Figure \ref{fig:circle}.

\subsection{Unstructured quadrilateral meshes}
\label{ssec:quads}
\begin{figure}[h!]
\begin{center}
\begin{subfigure}{.3\linewidth} \centering
    \includegraphics[width=\linewidth]{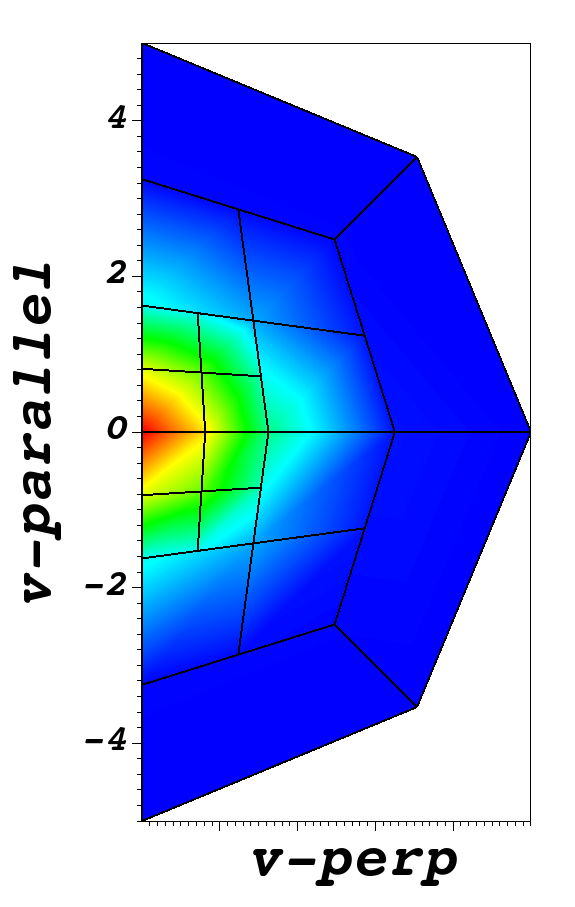}
    \caption{Q2}
%    \label{fig:enter-label}
\end{subfigure}
\begin{subfigure}{.3\linewidth} \centering
    \includegraphics[width=\linewidth]{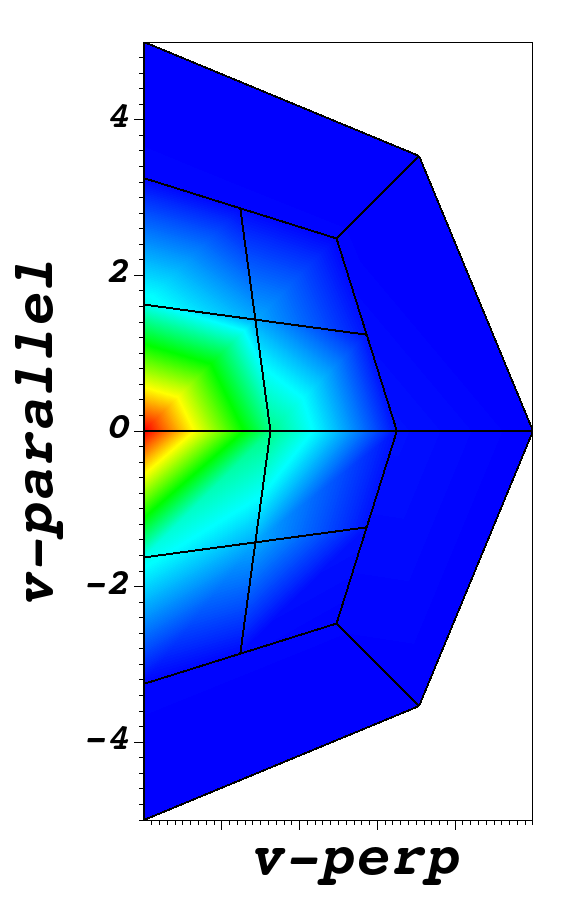}
    \caption{Q3}
%    \label{fig:enter-label}
\end{subfigure}
\begin{subfigure}{.3\linewidth} \centering
    \includegraphics[width=\linewidth]{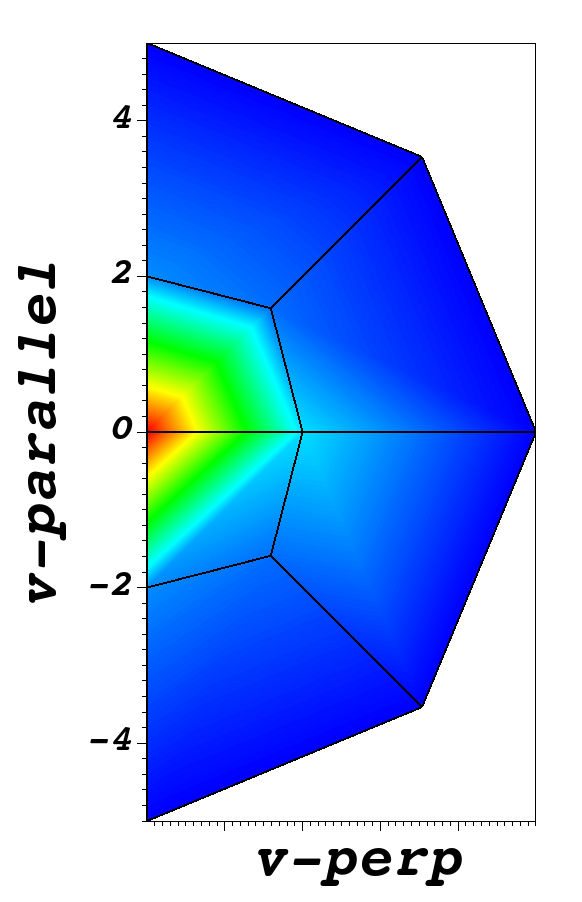}
    \caption{Q4}
%    \label{fig:enter-label}
\end{subfigure}
\caption{Semi-circular grids for electrons with tensor elements, from the $Q2$, $Q3$, $Q4$ test in \S\ref{sec:perfanisotropic} (linear interpolation in Visit results in visualization artifacts).}
\label{fig:circlequad}
\end{center}
\end{figure}

Tensor cell meshes have the advantage that our non-conforming AMR package, \textit{p4est}, can accommodate unstructured tensor cells, quadrilaterals and hexahedra, but not simplicies.
Nonconforming meshes can coarsen faster while maintaining well-shaped elements for better accuracy, which helps to reduce the number of cells required for a given level of accuracy. 
The disadvantage of tensor cells is that mesh generation is much less automated.
Velocity space distributions are, however, relatively simple in many cases with near-Maxwellian distributions.
The same adaptivity strategies developed for the block structured meshes apply to unstructured tensor cell meshes, such as refinement about the origin (eg, Figures \ref{fig:shifted_i0}) and the parallel axis (eg, Figures \ref{fig:shifted_e0}-\ref{fig:shifted_e4}).
Figure \ref{fig:circlequad} shows an example of a quadrilateral mesh used in the verification test in \S\ref{ssec:perf_aniosotropic_data}.

%In $2V$ the solves, are subdominant and in $3V$ the time is dominated by the Jacobian creation, which is expected because this the $\Order{N^2}$ work complexity algorithm.
%Note that the mass matrix creation is essentially only finite element assembly and sparse matrix assembly and thus the Jacobian time minus the Mass time is the cost of the Landau kernel.
%\clearpage
\section{Anisotropic relaxation verification test}
\label{sec:verify}

A plasma resistivity test was presented in previous work to verify the Landau collision operator and place constraints on the mesh sizes (\S IV.B in \cite{Adams2022a}).
Plasma resistivity does not test the collision rate, the $\nu_{\alpha\beta}$ term in (\ref{eq:landau1}), nor does it include anisotropic distribution functions.
An anisotropic relaxation test, where each species of a deuterium plasma is initialized with different parallel and perpendicular temperatures, was reported by Hager et al. \cite{Hager2016} and is the basis of this test.

\subsection{Model problem}

A deuterium plasma is initialized with an anisotropic distribution with respect to mean parallel and perpendicular speed within each species and between the two species, and evolved to equilibrium.
The modified Maxwellian distribution is defined with an anisotropic parameter $\alpha$ ($\alpha=1.3$ herein) as  
%$$f_s(v_\parallel,v_\perp)=\frac{n_s m_s^{3/2}}{\alpha( k_b T_{s,\parallel})^{3/2}}\exp{\left[-m_s\frac{(v_\parallel-u_s)^2+v_\perp^2/\alpha}{2eT_{s,\parallel}}\right]},$$
\begin{equation}
\label{eq:anisotropic-maxwellian}
\  f_s \left (v_\parallel,v_\perp\right ) = \frac{n_s}{\alpha} \left(\frac{1}{\pi\theta}\right)^{3/2} \exp [ -\frac{(v_\parallel-u_s)^2+v_\perp^2/\alpha}{\theta}],
\end{equation}
with $\theta=2k_bT_s/m_s v_0^2$
where $n_s$ is the number density of species $s$ ($n_s=10^{20}$ herein) and for the shifted case the shift $u_s = -1.5 \cdot sgn(s) \cdot m_0 / m_s$ with $sgn(s) = -1$ for electrons and $sgn(s) = 1$ for ions, which results in zero net parallel momentum and a significant shift of the electrons.
%The shifted case is shown for a qualitative demonstration of the collision physics, whereas the unshifted chase is presented with analytical results. 

The total temperature is defined as $T_s = (2 \cdot T_\perp + T_\parallel)/3$ and, in a finite element context, is computed with 
$$T_{s}\left (v_\parallel,v_\perp \right ) \equiv \frac{m_s v_0^2 \int d{{v}}{f_s\left( v_\perp,v_\parallel\right) \cdot  v_\perp \cdot \left(\left(v_\parallel - u_s\right)^2 + v_\perp^2\right)}}{3 \int d{ v } f_s\left(v_\perp,v_\parallel\right) \cdot  v_\perp}.$$
These temperatures are evaluated in MKS units in the code and scale by $\sim 6.24e\times 10^{18}$ to convert from Joules to electron volts (eV).
Likewise the parallel and perpendicular temperatures are computed according to 
$$T_{s,\perp}\left (v_\parallel,v_\perp \right ) \equiv \frac{m_s v_0^2 \int d{{v}}{f_s\left(v_\perp,v_\parallel\right) \cdot  v_\perp \cdot v_\perp^2}}{2\int d{ v }{f_s\left(v_\perp,v_\parallel\right) \cdot  v_\perp}}$$ 
and 
$$T_{s,\parallel}\left (v_\parallel,v_\perp \right ) \equiv \frac{m_s v_0^2 \int d{{v}}{f_s\left( v_\perp,v_\parallel\right) \cdot  v_\perp \cdot \left(v_\parallel - u_s\right)^2}}{\int d{ v } f_s\left(v_\perp,v_\parallel\right) \cdot  v_\perp}.$$
These definitions generate an electron parallel temperature $T_{e,\parallel} = 300$ eV, perpendicular temperature $T_{e,\perp} = 390$ eV, and ion parallel temperature $T_{i,\parallel} = 200$ eV, perpendicular temperature $T_{i,\perp} = 260$ eV, as reported by Hager et al. \cite{Hager2016}, and as can be seen at $t=0$ in the plots in Figure \ref{fig:temperature-history}.
Two cases of anisotropic initial conditions are investigated: 1) \textit{non-shifted} $u_s=0$ and 2) \textit{shifted} $u_e = 1.5, u_i = 1.5 (m_e/m_i),$ $ m_e/m_i \approx 1/3,671$.

\subsection{Comparison with analytical results}
\label{sec:anisotropic_verify}

The NRL Plasma Formulary provides formulas for inter and intra-species thermalization rates (Appendix \ref{sec:nrl}) that generate analytical temperature histories for comparison with the computed results.
The Plasma Formulary also provides formulas for Coulomb logarithms, which are used for both the computed and analytical temperature histories.
All three relevant Coulomb logarithms are between $6.8$ and $7.5$ in this study.

\begin{figure}[h!]
\begin{center}
    \begin{subfigure}{0.49\linewidth} \centering
        \includegraphics[width=\linewidth]{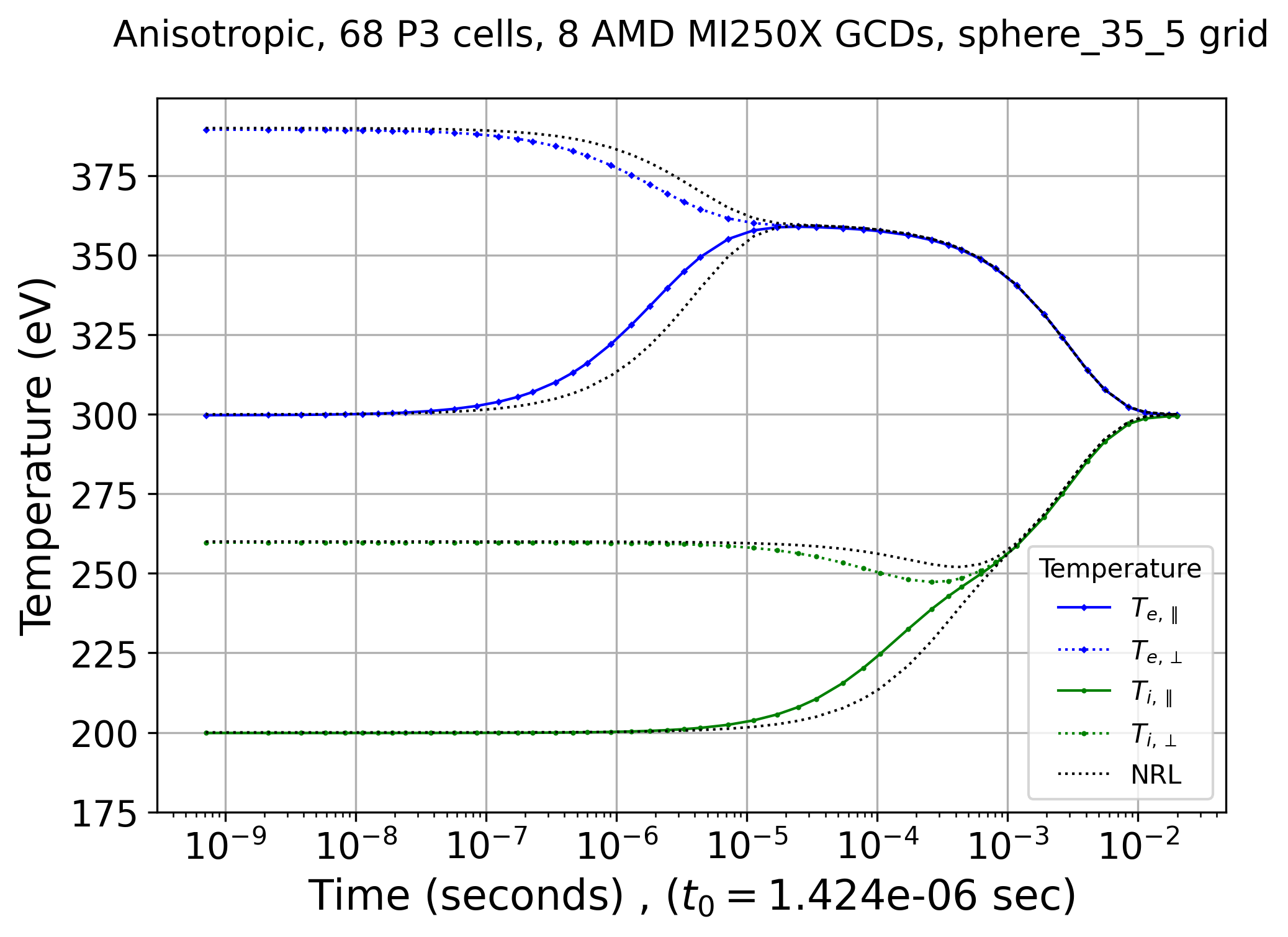}
       \caption{$P3$, 34 cell per species semi-circular mesh} \label{fig:temperature-max-Q4-1AMR}
    \end{subfigure}
    \begin{subfigure}{0.48\linewidth} \centering
        \includegraphics[width=\linewidth]{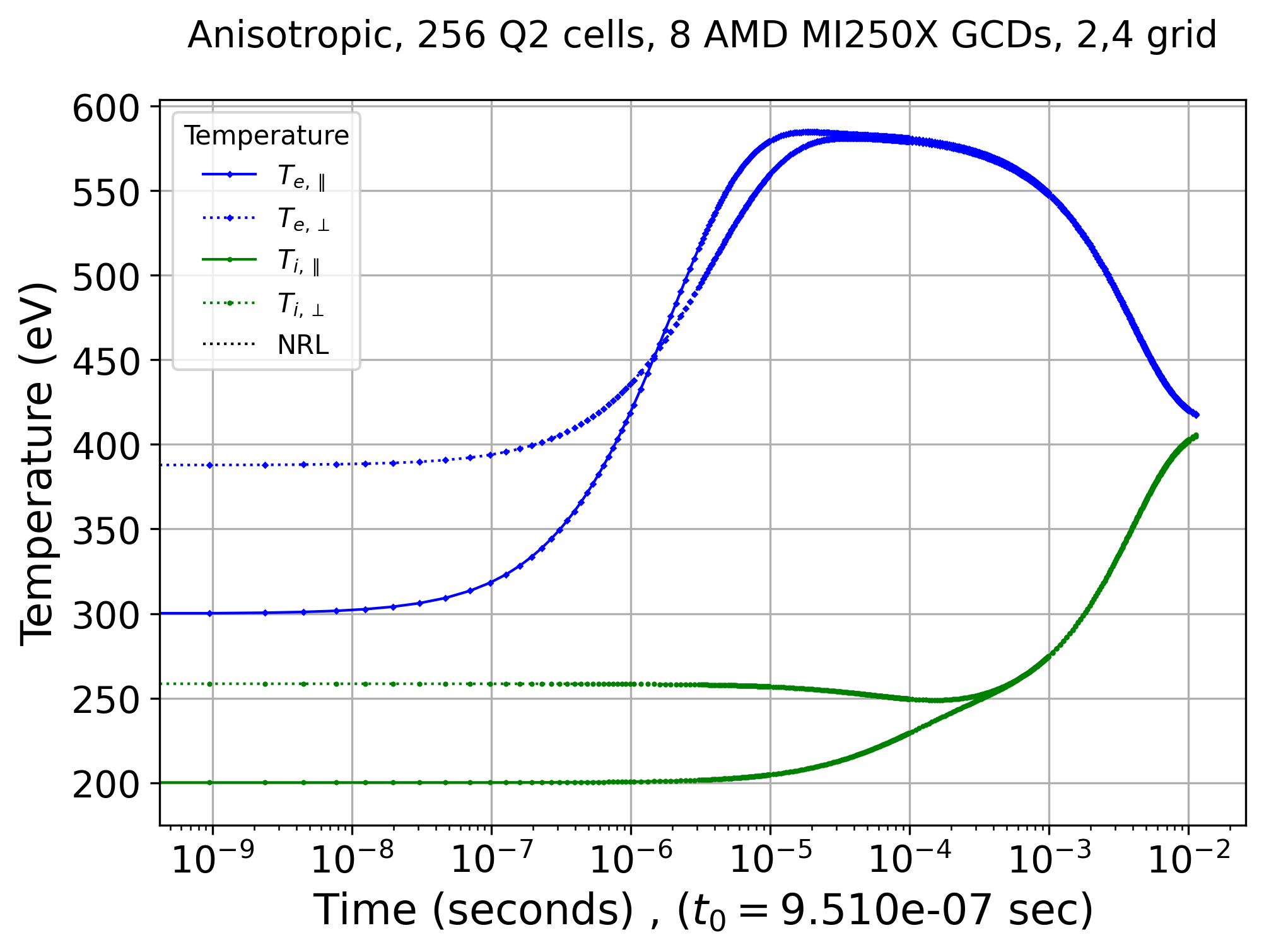}
        \caption{Shifted Bi-Maxwellian case} \label{fig:temperature-shifted}
    \end{subfigure}
\caption{Anisotropic relaxation test temperature vs. time of the $P3$ element case, plotted with an analytical NRL results for the non-shifted case (left), and the shifted Maxwellian case with a block-structured AMR grid and $Q2$ elements (right)}
\label{fig:temperature-history}
\end{center}
\end{figure}

Figure \ref{fig:temperature-max-Q4-1AMR} shows the temperatures as a function of time and the electron-electron collision period $t_0$ in the non-shifted cases, with $P3$ elements.
The discrepancy between the computed and NRL results is due to the fact that the NRL rates are based on Maxwellian distributions.
While this test thermalizes to a Maxwellian distribution, the plasma is significantly non-Maxwellian during the evolution.
Our results are in line with other published results \cite{Hager2016,Zonta2021,PusztayKnepleyAdams2022}.

\subsection{Qualitative observation of the shifted Maxwellian dynamics}

\begin{figure}[h!]
\begin{center}
    \begin{subfigure}{0.32\linewidth} \centering
        \includegraphics[width=\linewidth]{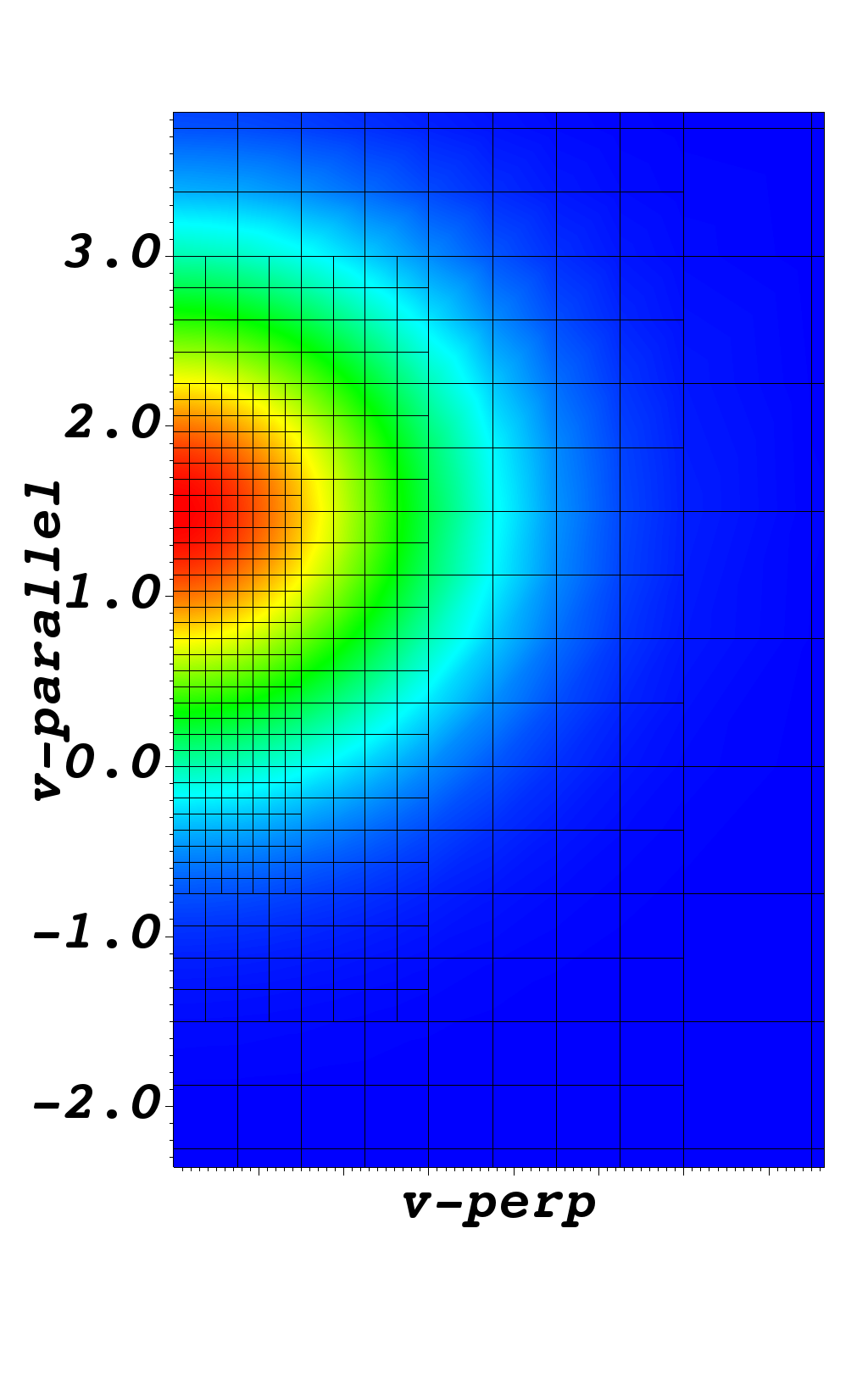}
        \caption{$t=0$ electrons} \label{fig:shifted_e0}
    \end{subfigure}
    \begin{subfigure}{0.32\linewidth} \centering
        \includegraphics[width=\linewidth]{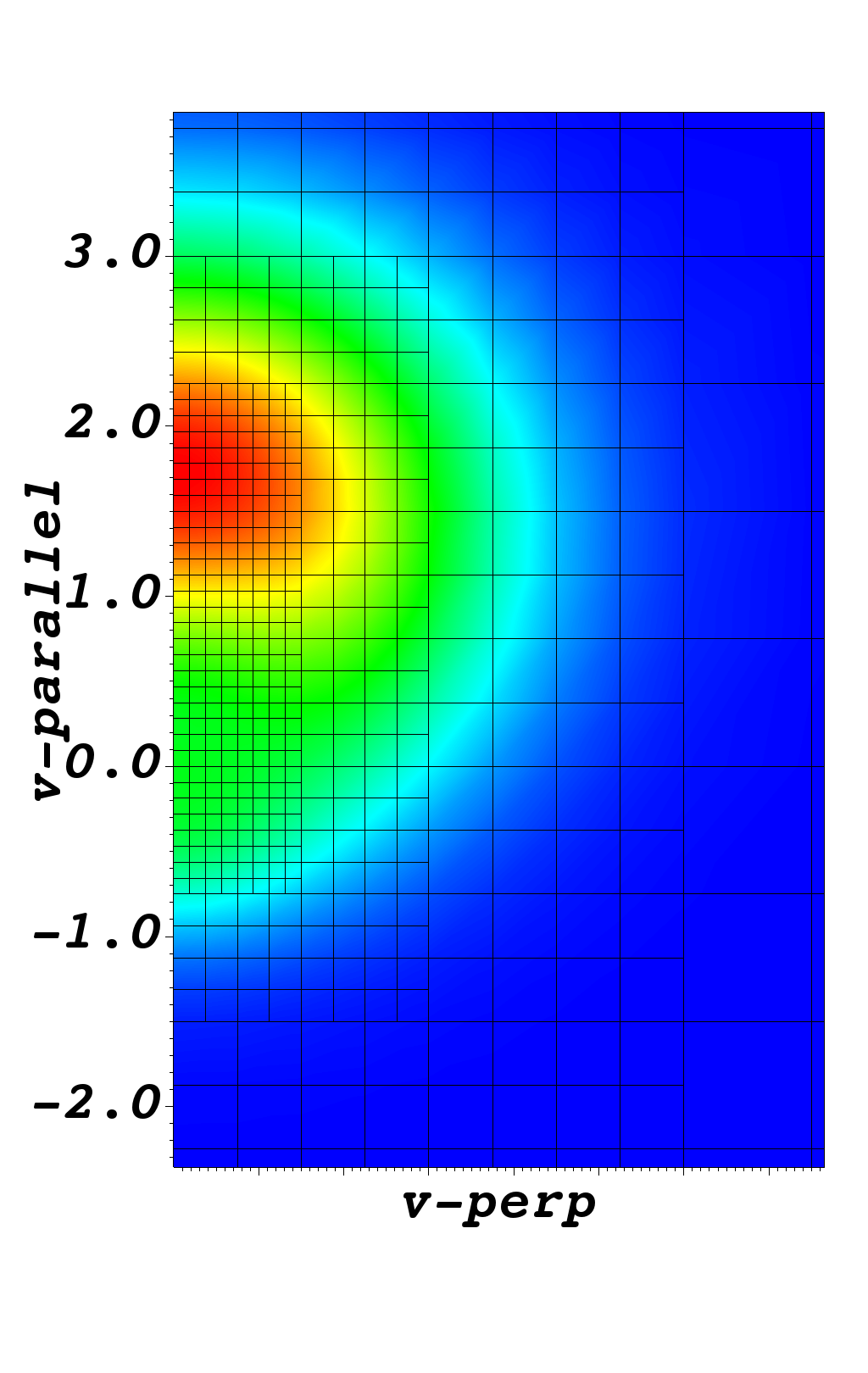}
        \caption{$t=0.33 \tau_e$} \label{fig:shifted_e1}
    \end{subfigure}
    \begin{subfigure}{0.32\linewidth} \centering
        \includegraphics[width=\linewidth]{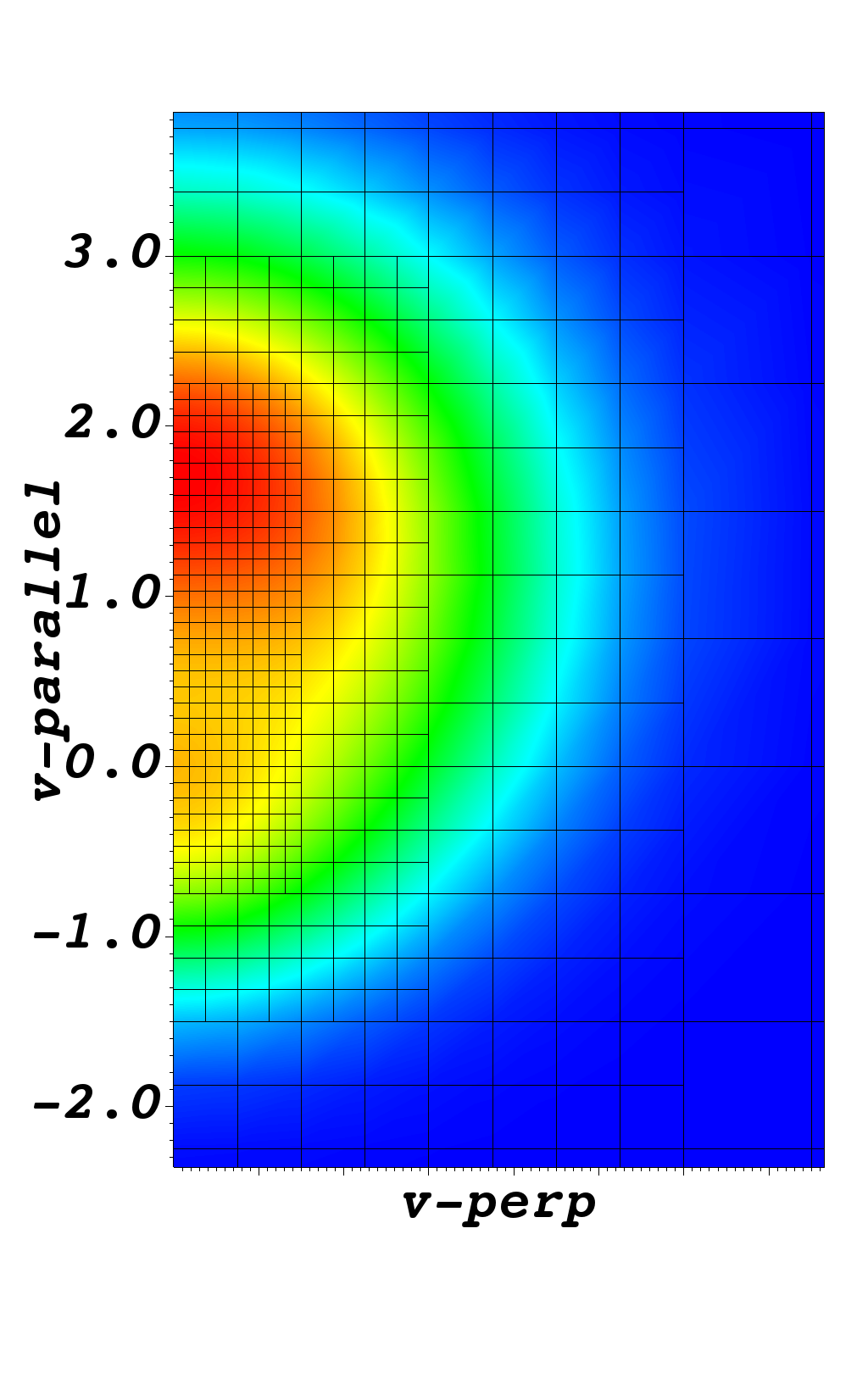}
        \caption{$t=1.37 \tau_e$, bi-modal} \label{fig:shifted_e2}
    \end{subfigure}\\
    \begin{subfigure}{0.32\linewidth} \centering
      \includegraphics[width=\linewidth]{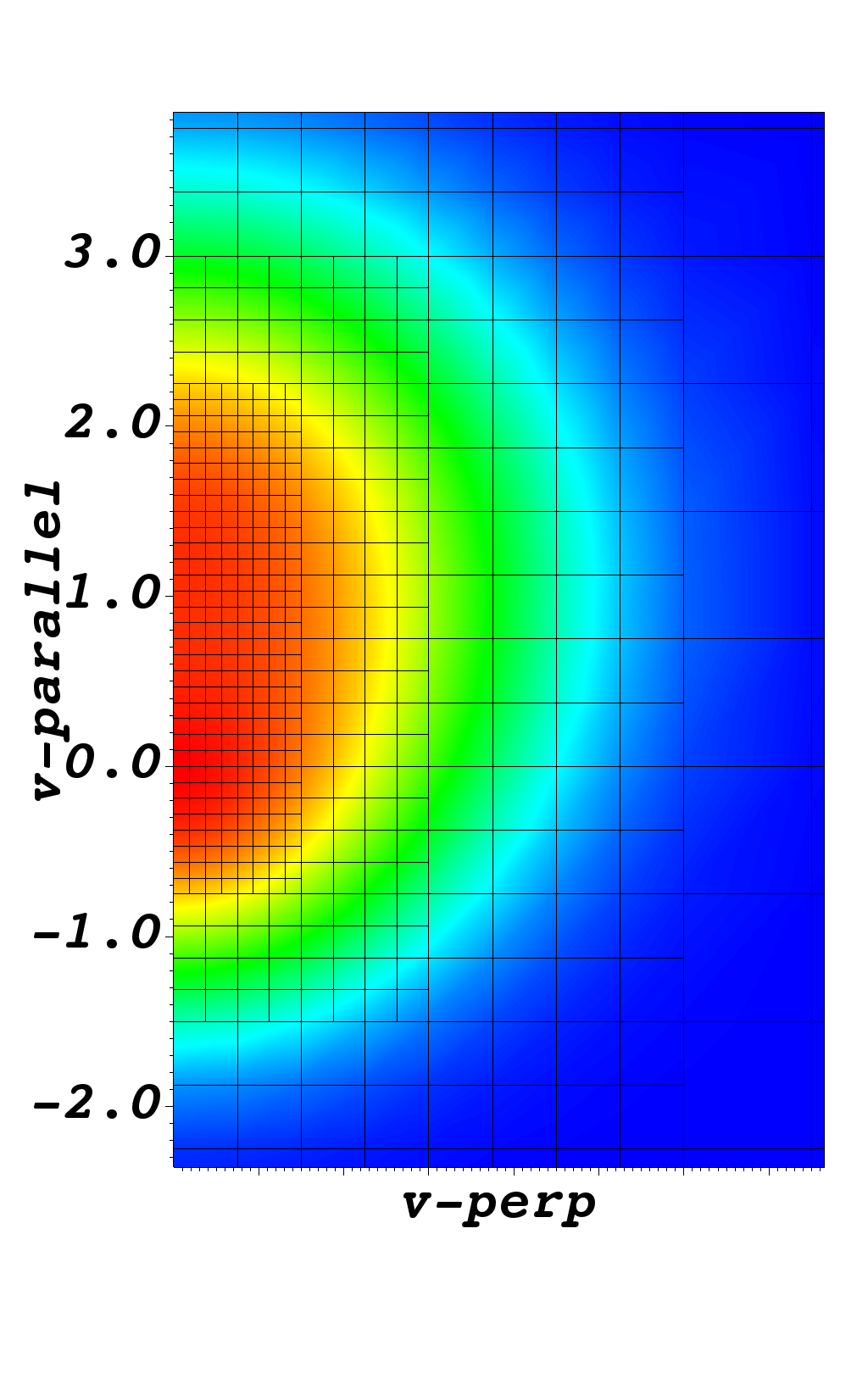}
      \caption{$t=2.49 \tau_e$ electrons}\label{fig:shifted_e3}
    \end{subfigure}
    \begin{subfigure}{0.32\linewidth} \centering
        \includegraphics[width=\linewidth]{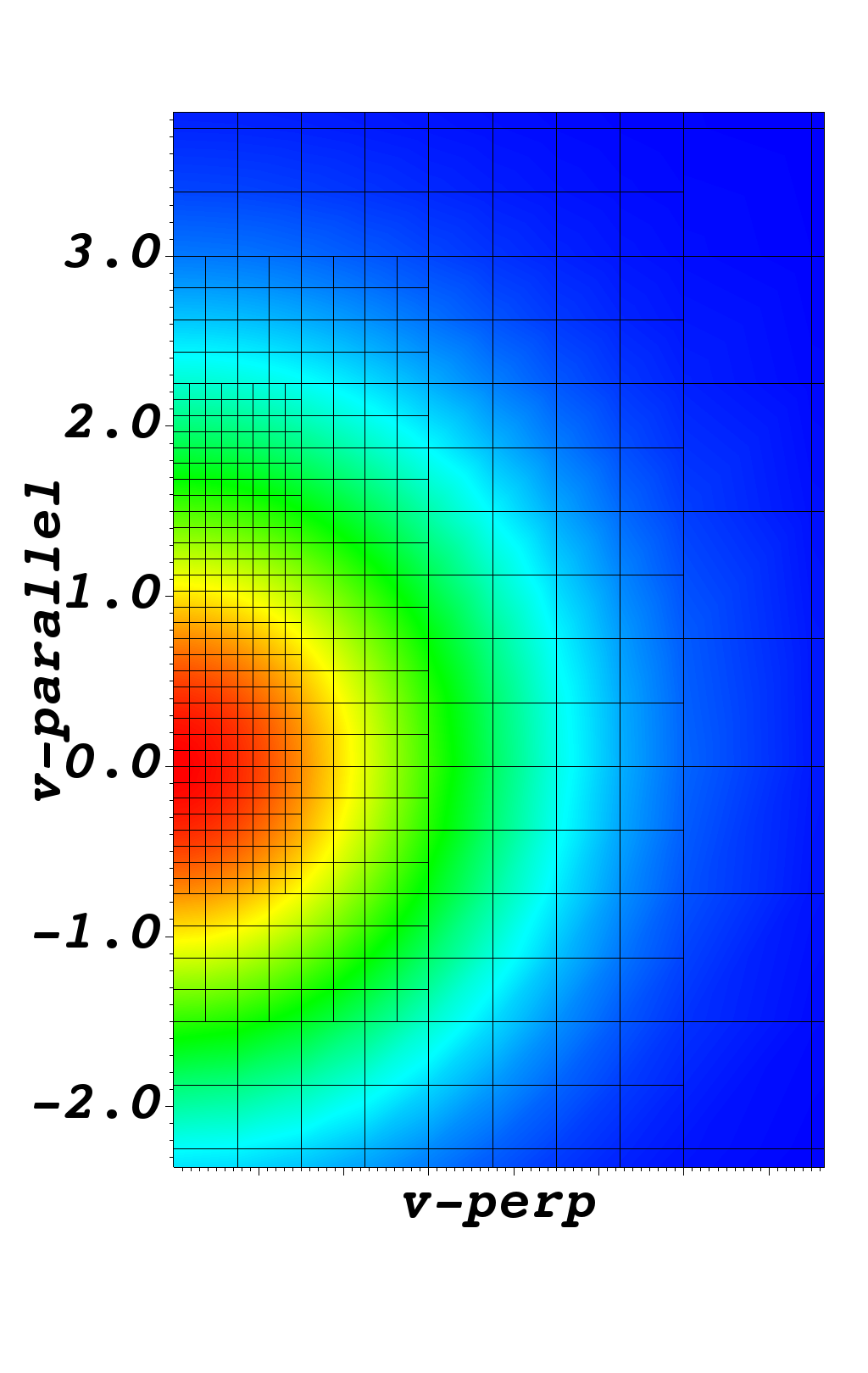}
        \caption{$t=23.1 \tau_e$} \label{fig:shifted_e4}
    \end{subfigure}
    \begin{subfigure}{0.34\linewidth} \centering
        \includegraphics[width=\linewidth]{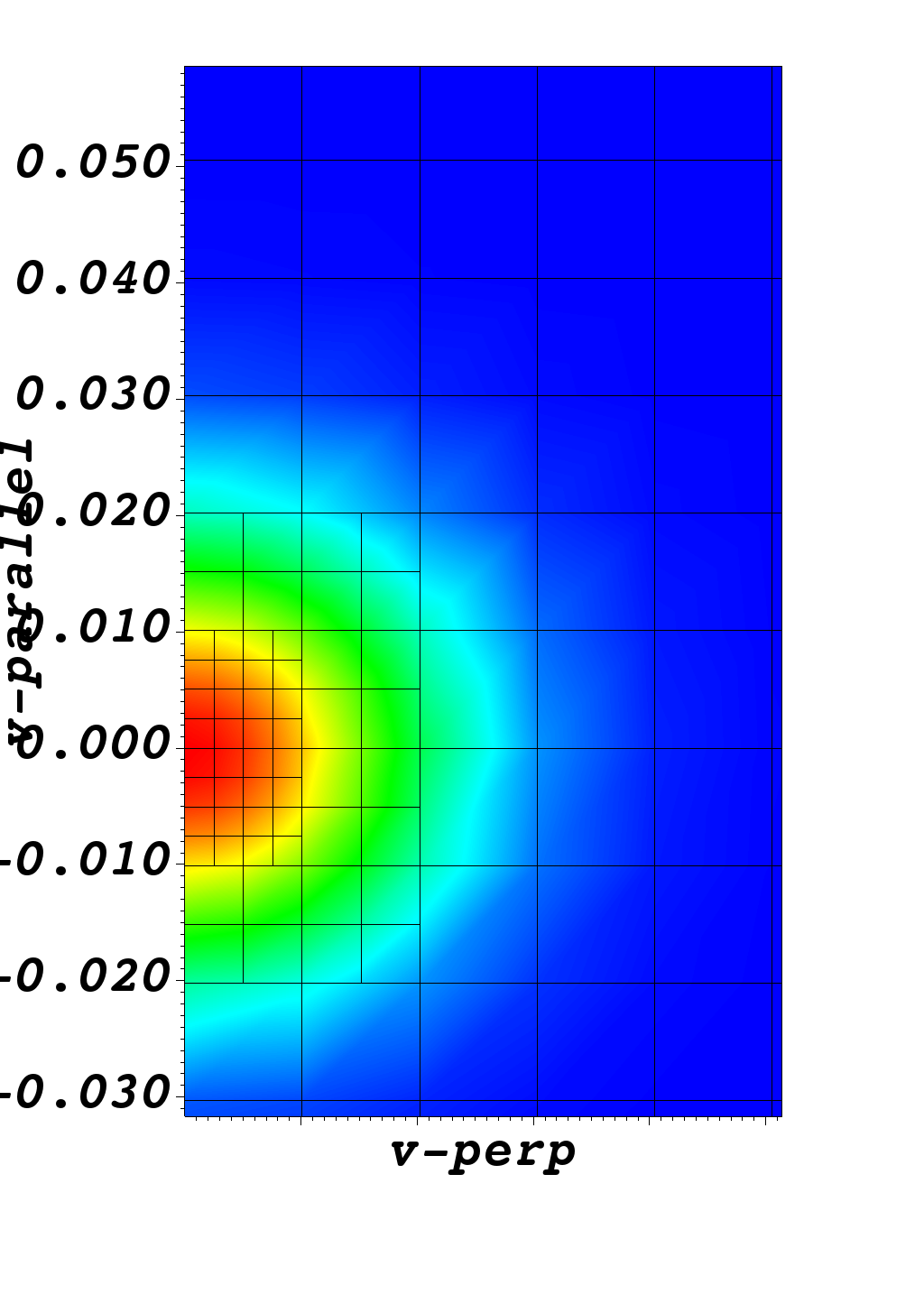}
        \caption{$t=0$, ions} \label{fig:shifted_i0}
    \end{subfigure}
\caption{Detail of the electron distributions, shifted Maxwellian deuterium plasma, ions near origin: (a) initial condition, (b) penumbra in shift and early Maxwellian population, (c) bi-modal distribution, (d) mid-thermalization, (e) near full thermalization, (e) initial condition of ions}
\label{fig:meshes}
\end{center}
\end{figure}

Figure \ref{fig:temperature-shifted} shows a solution to the shifted Maxwellian case, illustrated in Figure \ref{fig:meshes}, where the kinetic energy from the parallel flow is first transmitted to the parallel temperature of the electrons via collisions, followed by the isotropization of the electrons, and then the ion isotropization and the complete thermalization of the plasma.
Note, the kinetic energy in the parallel flow or drift velocity is not used in the temperature calculation, resulting in an increase temperature as this case thermalizes.

The dynamics of the shifted Maxwellian thermalization is of interest in developing an intuitive understanding of the physics of collisions.
%This section examines the shifted Maxwellian dynamics.
%The visualization software uses linear interpolation that distorts the codes data and a highly resolved AMR mesh is used to aid visualization.
Figure \ref{fig:meshes} shows a detail of the electron distribution function for several early times in the thermalization of the shifted problem on a highly refined mesh using block-structured mesh adaptivity.
Figures \ref{fig:shifted_e0}-\ref{fig:shifted_e2} show the initial thermalization of electrons, followed by the shift of the bulk of the electrons to eventual (near) thermalization in Figures \ref{fig:shifted_e3}-\ref{fig:shifted_e4}.
Figure \ref{fig:shifted_i0} shows the initial ion distribution (note, electron and ion grids have different topology and scaling).
The times of each figure refers to the temperature histories, normalized with $t_0$, shown in Figure \ref{fig:temperature-shifted}.

\section{Performance results for the anisotropic relaxation test}
\label{sec:perfanisotropic}

This section investigates the performance of the anisotropic relaxation test with the two single node configurations in \S\ref{sec:throughput_perf}, using the unstructured grids shown in Figure \ref{fig:circle}, with $P2-P4$ triangle elements, and the quadrilateral meshes in Figure \ref{fig:circlequad} with $Q2-Q4$ elements.
%An error metric is required to understand costs.
%A single scalar metric is challenging because 
The domain radius is optimized by hand for the $P2$ grid and uses a domain size of $5.5$, but otherwise a radius of $5.0$ is used.
The quadrilateral meshes are optimized to some extent, although $Q2$ elements are not served well with the current meshing.

A scalar error metric is desired to understand costs versus accuracy and to constrain the degree of mesh refinement.
To this end, error is defined as the difference in any of the four measured temperatures from the correct value of $300$ eV at the converged state: $error \equiv \max_{s\in \left[e,i\right], d \in \left[\parallel,\perp\right]} \lvert T_{s,d} - 300 \rvert$, and is listed in Tables \ref{tab:timings-NVIDIAP}, \ref{tab:timings-AMDP} and \ref{tab:timings-NVIDIAQ}.
This is not a rigorous metric because the two species grids are nondimensionalized with the initial parallel thermal speed of the single species on the grid, which are different.
While both species have the same thermal speed and (Maxwellian) distribution at equilibrium, different computational domains results in different discretization errors in the temperature integrals.
We observe decent correlation of this metric and refinement, but these results should be verified with a qualitative assessment of the temperature histories compared to the NRL data (see Appendix \ref{ap:details}).
We observe a decent correlation between this metric and the qualitative assessment of the temperature histories.

\subsection{Solver stack}
\label{sec:anisotropic_ts}

An adaptive three-stage Runge-Kutta time integrator with an initial time step of $0.001$, in units of the electron-electron collision time $\tau_e$, is used to  capture the early dynamics accurately.
The time step quickly increases to an imposed maximum value of $1.0$ as the plasma reaches equilibrium.
Each of the three stages uses a nonlinear solve with a relative residual tolerance of $10^{-14}$ and all three moments of interest, density, parallel momentum, and energy, are conserved to this scale in each time step. 
The total simulation ($T = 14,000 \tau_e$ and $>14,000$ time steps) conserves energy to about 12 digits.
%Quadratic square elements (Q2) are used throughout unless otherwise noted.
%The batched TFQMR solver in PETSc, written in Kokkos, with Jacobi preconditioning is used.
\subsection{Anisotropic relaxation test performance data}
\label{ssec:perf_aniosotropic_data}

%The NVIDIA A100 and AMD MI250X single node configurations in \S\ref{sec:throughput_perf} are used to investigate performance of the full anisotropic relaxation test simulation time.
The AMD MI250X results use ROCm 5.2 and on the \textit{Frontier} machine and the NVIDIA A100 results use the \textit{Perlmutter} machine.
Both CUDA and HIP allow for specifying launch bounds at compile time, defined as {\tt <maxThreadsPerBlock, minBlocksPerMultiprocessor>} and {\tt <maxThreadsPerBlock, minWarpsPerEU>} respectively.
We found the default launch bounds of $<1024,1>$ were not ideal, especially for the MI250X.
We found that $<256,1>$ was reasonable for the MI250X and that $<256,2>$ was reasonable for the A100.
A batch size of 128 on AMD with 8 MPI processes and 256 on NVIDIA with 4 MPI processes is used, resulting in 1,024 full anisotropic relaxation problems running simultaneously.
These 1,024 problems are identical and mimic an application setting where multiple spatial points are processed simultaneously.

%With 128 or 64 Kokkos thread groups, each of which which corresponds to NVIDIA Streaming Multiprocessor (SM) or AMD Compute Units on NVIDIA and AMD, respectively, 
%The linear solvers do not saturate the nodes with batch sizes and throughput can be increase by increasing the batch size in these experiments.
%However, this batch size is warranted for economy, and the solver times are subdominant and would only become more so with the larger batch size.

Three simplex cases are investigated, $P2,P3$ and $P4$ elements with the grids shown in Figure \ref{fig:circle}, as well as three quadrilateral cases, $Q2,Q3$ and $Q4$ elements with the grids in Figure \ref{fig:circlequad}.
Tables \ref{tab:timings-NVIDIAP}, \ref{tab:timings-AMDP} and \ref{tab:timings-NVIDIAQ} list the number of cells, the number of integration points (IPs), the number of equations in the matrix for each problem (with two species, each linear system is one half this size), the average number of non-zeros per row, the order $P$ or $Q$ of the finite elements, the  matrix construction and linear solver times, the total run time, and the percent error. 
\begin{table}[h!]
\caption{Anisotropic thermalization timings (seconds) on 4 NVIDIA A100 GPUs}
\label{tab:timings-NVIDIAP}
\resizebox{\columnwidth}{!}{%
\begin{tabular}{cccccrrrr}
\toprule
\# cells & \# IPs & \# eqs & nnz/row & P & Jacobian & Solve & Total time & error (\%) \\
\midrule
144 & 864 & 318 & 10.8 & 2 & {\cellcolor[HTML]{CE2827}} \color[HTML]{F1F1F1} 4,930 & {\cellcolor[HTML]{15904C}} \color[HTML]{F1F1F1} 424 & {\cellcolor[HTML]{CE2827}} \color[HTML]{F1F1F1} 5,424 & {\cellcolor[HTML]{FDAD60}} \color[HTML]{000000} 0.67 \\
68 & 816 & 350 & 15.5 & 3 & {\cellcolor[HTML]{DC3B2C}} \color[HTML]{F1F1F1} 4,785 & {\cellcolor[HTML]{FEE797}} \color[HTML]{000000} 495 & {\cellcolor[HTML]{D62F27}} \color[HTML]{F1F1F1} 5,352 & {\cellcolor[HTML]{15904C}} \color[HTML]{F1F1F1} 0.18 \\
24 & 384 & 226 & 20.8 & 4 & {\cellcolor[HTML]{15904C}} \color[HTML]{F1F1F1} 1,376 & {\cellcolor[HTML]{CE2827}} \color[HTML]{F1F1F1} 544 & {\cellcolor[HTML]{15904C}} \color[HTML]{F1F1F1} 2,009 & {\cellcolor[HTML]{CE2827}} \color[HTML]{F1F1F1} 0.84 \\
\bottomrule
\end{tabular}
}
\end{table}
\begin{table}[h!]
\caption{Anisotropic thermalization timings (seconds) on 8 AMD MI250X GCDs}
\label{tab:timings-AMDP}
\resizebox{\columnwidth}{!}{%
\begin{tabular}{cccccrrrr}
\toprule
\# cells & \# IPs & \# eqs & nnz/row & P & Jacobian & Solve & Total time & error (\%) \\
\midrule
144 & 864 & 318 & 10.8 & 2 & {\cellcolor[HTML]{CE2827}} \color[HTML]{F1F1F1} 3,948 & {\cellcolor[HTML]{15904C}} \color[HTML]{F1F1F1} 660 & {\cellcolor[HTML]{CE2827}} \color[HTML]{F1F1F1} 4,746 & {\cellcolor[HTML]{FDAD60}} \color[HTML]{000000} 0.67 \\
68 & 816 & 350 & 15.5 & 3 & {\cellcolor[HTML]{F47044}} \color[HTML]{F1F1F1} 3,510 & {\cellcolor[HTML]{FEE999}} \color[HTML]{000000} 826 & {\cellcolor[HTML]{EE613E}} \color[HTML]{F1F1F1} 4,421 & {\cellcolor[HTML]{15904C}} \color[HTML]{F1F1F1} 0.18 \\
24 & 384 & 226 & 20.8 & 4 & {\cellcolor[HTML]{15904C}} \color[HTML]{F1F1F1} 959 & {\cellcolor[HTML]{CE2827}} \color[HTML]{F1F1F1} 943 & {\cellcolor[HTML]{15904C}} \color[HTML]{F1F1F1} 1,982 & {\cellcolor[HTML]{CE2827}} \color[HTML]{F1F1F1} 0.84 \\
\bottomrule
\end{tabular}
}
\end{table}

\begin{table}[h!]
\caption{Anisotropic times (seconds) using tensor elements on the NVIDIA node}
\label{tab:timings-NVIDIAQ}
\resizebox{\columnwidth}{!}{%
\begin{tabular}{cccccrrrr}
\toprule
\# cells & \# IPs & \# eqs & nnz/row & Q & Jacobian & Solve & Total time & error (\%) \\
\midrule
36 & 324 & 154 & 14.3 & 2 & {\cellcolor[HTML]{15904C}} \color[HTML]{F1F1F1} 1,189 & {\cellcolor[HTML]{15904C}} \color[HTML]{F1F1F1} 236 & {\cellcolor[HTML]{15904C}} \color[HTML]{F1F1F1} 1,543 & {\cellcolor[HTML]{CE2827}} \color[HTML]{F1F1F1} 2.83 \\
24 & 384 & 236 & 22.7 & 3 & {\cellcolor[HTML]{CE2827}} \color[HTML]{F1F1F1} 1,447 & {\cellcolor[HTML]{E5F49B}} \color[HTML]{000000} 328 & {\cellcolor[HTML]{F36B42}} \color[HTML]{F1F1F1} 1,897 & {\cellcolor[HTML]{30A356}} \color[HTML]{F1F1F1} 0.77 \\
12 & 300 & 226 & 31.4 & 4 & {\cellcolor[HTML]{FDB567}} \color[HTML]{000000} 1,375 & {\cellcolor[HTML]{CE2827}} \color[HTML]{F1F1F1} 456 & {\cellcolor[HTML]{CE2827}} \color[HTML]{F1F1F1} 1,953 & {\cellcolor[HTML]{15904C}} \color[HTML]{F1F1F1} 0.64 \\
\bottomrule
\end{tabular}
}
\end{table}

This data shows that the A100 and MI250X performance is comparable with the solver running faster on the A100 and the matrix construction running faster on the MI250X node.
These times correlate with the accuracy with $P4$ being the fastest, by far, and $P3$ being the most accurate on the simplex grids, and the quadrilateral grids being a bit more efficient than the simplex grids with the exception of the $Q2$ case. % that is clearly not a good mesh.
%However, $P2$ is both slower and less accurate than $P3$ and thus not a desirable option for this problem.

%node in the Jacobian matrix construction and slightly faster in the linear solver.
%Note, the MI250X node was observed to be faster for the linear solver in Tables \ref{tab:tfqmr-Perlmutter-solve_throughput} and \ref{tab:tfqmr-Crusher-solve_throughput}, which use smaller grids and larger batch sizes.

\section{Hardware utilization}
\label{sec:utilization}

This section analyses the efficacy of the hardware utilization in the results in \S\ref{sec:throughput_perf} and \S\ref{sec:perfanisotropic}.
The throughput performance study in \S\ref{sec:throughput_perf} is analysed on the A100 with NVIDIA's Nsight Systems in \S\ref{sec:nvidia_hw} and the anisotropic relaxation test in \S\ref{sec:perfanisotropic} is analysed on the AMD MI250X with the TAU performance tools in \S\ref{sec:amd_hw} \cite{tau}.
One of the four A100 GPUs and one of the eight MI205X GCDs are used with one MPI process on our nodes for analysis.

PETSc supports the use of these two architectures with both CUDA and HIP back-ends as well as a \textit{Kokkos Kernels} back-end that supports both architectures.
The \textit{Kokkos Kernels} back-end, using the built-in \textit{Kokkos} numerical kernels, is used for all performance results herein.
The codes used for these tests are examples in PETSc.
Instructions for reproducing these results, the raw data, and the scripts that generate the plots are publicly available (Appendix \ref{sec:ad}).

The NVIDIA's Nsight Systems data in \S\ref{sec:throughput_perf} shows that $97\%$ of the total run time is spent on the GPU and $100\%$ flops in the full collision time advance are executed on the device, after the grid setup phase.
The linear solver and Jacobian construction kernels are written in \textit{Kokkos}, the nonlinear solver and time integrator use the appropriate \textit{Kokkos Kernels} sparse matrix a vector back-ends.

\subsection{NVIDIA hardware utilization}
\label{sec:nvidia_hw}

%\begin{figure}[htbp]
%\begin{center}
%\includegraphics[width=1.\linewidth]{nsys-2d-NewtIt.png}
%\caption{Nsight Systems view of a typical Newton iteration with: CUDA device ``97.0\% Kernels" (dark blue, middle row) with three large kernels (left to right): the Jacobian construction, mass matrix construction and linear solve (``batch-kokkos-solve"). The Jacobian is proceeded by a kernel that builds the function values and derivatives at the integration points and each matrix method is followed by COO matrix assembly kernel.}
%\label{fig:nsys-2D}
%\end{center}
%\end{figure}

The analysis of the hardware utilization in the A100 kernel is divided into the analysis of the Jacobian matrix and the mass matrix construction, and the batch solver in $2V$ and $3V$ on the experiments in \S\ref{sec:throughput_perf}.
The NVIDIA Nsight Compute tool is used to gather several hardware metrics from the largest batch size in Tables \ref{tab:parts-Perlmutter-2V} and \ref{tab:parts-Perlmutter-3V}.
Table \ref{tab:rooflinedata} presents some of the raw Nsight Compute data.

\begin{table}[htbp]
\centering
\caption{Nsight Compute data: Jacobian (Jac), Mass (M), Solver (Sol)}
\label{tab:rooflinedata}
\begin{tabular}{lrrrrrr}
\toprule
Data &  Jac-2V &  M-2V &  Sol-2V &  Jac-3V &  M-3V &  Sol-3V \\
\midrule
DRAM (GB/s)    &   75.80 &  1230 &   28.18 &   38.33 &   946 &     538 \\
L1 (TB/s)      &    1.92 &  3.58 &    1.43 &    1.92 &  2.39 &    1.59 \\
L2 (GB/s)      &     747 &  4010 &     881 &     266 &  2810 &    1870 \\
dadd/cycle     &     163 &   155 &     156 &   76.20 & 91.80 &   35.12 \\
dfma/cycle     &    1155 &     0 &     168 &     546 &     0 &   36.11 \\
dmul/cycle     &     526 &   329 &   64.50 &     305 &   198 &    3.14 \\
TFlop/sec      &    4.23 &  0.68 &    0.89 &    2.06 &  0.41 &    0.16 \\
AI-L1          &    2.20 &  0.19 &    0.50 &    1.07 &  0.17 &    0.10 \\
Roofline-L1 \% &   43.60 & 18.27 &    9.18 &   21.27 & 12.19 &    8.11 \\
AI-L2          &    5.66 &  0.17 &    0.72 &    7.75 &  0.15 &    0.08 \\
Roofline-L2 \% &   43.60 & 54.20 &   16.70 &   21.27 &    38 &   25.30 \\
AI-DRAM        &   55.80 &  0.56 &   23.60 &   53.80 &  0.43 &    0.29 \\
R.F.-DRAM \%   &   43.60 & 63.60 &    9.18 &   21.30 & 48.90 &   27.80 \\
\bottomrule
\end{tabular}
\end{table}

Some points can be seen in this data.
\begin{itemize}
    \item The Jacobian kernel, with a high arithmetic intensity (AI) of $55.8$ with respect to DRAM memory movement in $2V$, is not a simple loop of fused multiply add (FMA) instructions as can be seen from lines 4-6 with only $62\%$ of the flops in FMA instructions. This limits the achievable percent of theoretical peak for this algorithm.
    \item The flop rate (line 7) is about $2x$ higher in $2V$ than $3V$. This is at least partially due to the Landau kernel $\mathbf U$ in (\ref{eq:landau1}) being more complex with a higher AI in $2V$, but this requires futher investigation.
    \item There are few flops and no FMAs in the mass matrix as this is essentially all assembly.
    \item The solver AI-DRAM is very high in $2V$ (23.6) and low in $3V$ (0.29). The theoretical AI of the solver (no cache) is about $\frac{1}{6}$. This data indicates that the solves are fitting in cache well in $2V$ but not at all in $3V$.
\end{itemize}

Tables \ref{tab:Perlmutter-bottlenecks} and \ref{tab:Perlmutter-notes} tabulate conclusions and notes from the Nsight Compute data.
\begin{table}[htbp]
  \centering
  \footnotesize
    \caption{Nsight Compute Bottlenecks}
    \label{tab:Perlmutter-bottlenecks}
    \begin{tabularx}{\textwidth}{llll}    %\ begin{tabular}{llllll}
    \toprule
Jacobian-2V &  Mass-2V &  Solve-2V \\ % &  Jacobian-3V &  Mass-3V &  Solve-3V \\
\midrule
FP64 pipe (57\%) &  L2 (70\%),  & L1 and instruction latency bound:  &  FP64 pipe (31\%),  \\ % &   L2 (50\%),  &   L2 (28\%), \\
                 &  DRAM (64\%) & L1 (43\%) instruction issue (39\%) &  L1 (24\%)          \\ % &   DRAM (49\%) &   DRAM (23\%) \\
    \bottomrule
    \end{tabularx}
\end{table}

\begin{table}[h!]
\centering
\footnotesize
\caption{Nsight Compute Notes}
\label{tab:Perlmutter-notes}
\begin{tabularx}{\textwidth}{lll}
\toprule
Jacobian-2V &  Mass-2V &  Jacobian-3V \\ % &  Mass-3V &  Solve-3V \\
\midrule
Roofline lower than           & low roofline peak b/c 1) low pipe      & Low pipe utilization \\ % & &  Utilization not higher         &  Memory   \\
FP64 pipe utilization         & utilization due to being L1 latency    & due to L1 latency    \\ % & &  partly due to load imbalance:  &  latency   \\
b/c DFMA instruction is       & bound. 2) instruction dominated by     & bound                \\ % & &  Theoretical occupancy 44\%,    &  bound     \\
62\% of all FP64 instructions & branch and integers. FP64 instructions &                      \\ % & &  achieved occupancy 34\%        &         \\
                              & $\approx 10$\% of total instructions   &                      \\ % & &                                 &         \\
\bottomrule
\end{tabularx}
\end{table}

\subsection{AMD hardware utilization}
\label{sec:amd_hw}

The analysis of the hardware utilization of the AMD MI250X uses the anisotropic relaxation test in \S\ref{sec:perfanisotropic}, which has two grids with one species each ($S=1,1$), as opposed to ten species and three grids ($S=1,1,8$) in NVIDIA model problems.
These two tests have different complexity profiles in the Jacobian kernel (Algorithm 1, \cite{Adams2022a}), with the the substantial work complexity and low memory complexity, in $2V$, of the Landau tensor (\cite{Hirvijoki2016}, Appendix A) being amortized in multi-species per grid problem.
The inner loop accesses data proportional to $S$, for the field and gradient field data, imposing more register pressure in the eight species per grid case, and the arithmetic intensity is nominally higher in the eight species case with reuse of this field data.
The individual grids are slightly larger in the anisotropic relaxation test used for the AMD studies, resulting in lower batch size requirements to saturate the hardware.
Both studies increase the batch size to observe the saturation of the hardware.
%The analysis focuses the analysis of the Jacobian matrix construction and the linear solver.

%\begingroup
%\let\clearpage\relax
%\include{frontier-data-2v-and-3v-data}
%\include{frontier-data-2v-data}
%\endgroup
\begin{table}[h!]
\centering
\footnotesize
\caption{Rocprof / TAU data: Jacobian (Jac), Mass (M), Solver (Sol).}
\label{tab:Frontier-notes}
\begin{tabularx}{0.675\textwidth}{lrrr}
\toprule
Data                           & Jac-2V & M-2V  & Sol-2V \\
\midrule
Workgroups                     & 256    & 256   & 256    \\
LDS usage                      & 512    & 16896 & 512    \\
Scratch usage                  & 248    & 176   & 0      \\
Vector registers used          & 120    & 56    & 48     \\
Scalar registers used          & 120    & 120   & 112    \\
\midrule
Effective LDS BW GB/s          & 744    & 764   & 1638   \\
Effective vL1D BW GB/s         & 3814   & 4640  & 2761   \\
Effective L2 BW GB/s           & 216    & 722   & 393    \\
Effective HBM BW GB/s          & 107    & 622   & 188    \\
\midrule
AI-vL1D                        & $<0.005$   & $<0.005$  & $<0.005$   \\
AI-L2                          & 0.03   & $<0.005$  & $<0.005$   \\
AI-HBM                         & 0.07   & $<0.005$  & $<0.005$   \\
TFLOP/s                        & 3.85   & 0.50  & 0.40   \\
\midrule
L2 Cache Hit \%                & 77.07  & 57.22 & 74.32  \\
vL1D Cache Hit \%              & 94.33  & 84.83 & 85.77  \\
\midrule
Vector Inst. \%                & 74.63  & 52.96 & 55.56  \\
Scalar Inst. \%                & 14.82  & 22.61 & 18.49  \\
Branch Inst. \%                & 2.05   & 4.34  & 7.28   \\
Vector Memory Inst. \%         & 4.33   & 10.83 & 4.15   \\
LDS Inst. \%                   & 1.22   & 2.20  & 5.15   \\
\midrule
Avg. Dependency Wait Cycles \% & 55.22  & 72.8  & 81.86  \\
Avg. Issue Wait Cycles \%      & 10.60  & 14.66 & 1.77   \\
Avg. Active Issue Cycles \%    & 37.18  & 12.54 & 16.37  \\
Avg. Active Wavefronts         & 58.07  & 55.78 & 55.82  \\
\bottomrule
\end{tabularx}
\end{table}

The AMD nodes have eight GCDs, each with 110 Compute Units that are each similar to the 108 Streaming Multiprocessors on NVIDIA A100.
As with the A100, the analysis of the hardware utilization in the MI250X kernel is divided into the analysis of the Jacobian matrix and the mass matrix construction.
%, but only the batch solver in $2V$ on the experiments in \S\ref{sec:throughput_perf}.
The TAU Performance System and Rocprof tools are used to gather several hardware metrics from the $P4$ test in Table \ref{tab:timings-AMDP} and a batch size of 64.
Table \ref{tab:Frontier-notes} presents some of the raw Rocprof data, post-processed with a Python script written with guidance from AMD engineers for interpreting the metrics.

Because the set of available AMD metrics differ from that provided by NVIDIA, different points can be seen in this data, but it is still a useful comparison.
\begin{itemize}
    \item Both as an absolute value and relative to the A100 results, the Jacobian kernel has low arithmetic intensity (AI-HBM, computed as total FLOP / effective bandwidth) of $0.07$ with respect to DRAM memory movement in $2V$. The kernel is apparently suffering from the cost of the relatively smaller L1 cache of the MI250X, relatively low L2 hit rate, and register spilling to scratch despite the application of launch bounds. This limits the achievable performance for this algorithm.
    \item Despite the lower algorithmic intensity, the flop rate (TFLOP/s) for all three kernels is still comparable to the performance on the A100, although slightly lower.
    \item The solver AI-DRAM is low in $2V$ ($<0.005$) relative to the A100 data. This indicates that the solves are not fitting in cache in $2V$. The number of nonzeros in each of the linear systems is similar in this $P4$ unstructured test and the $Q3$ semi-structured grids of the NVIDIA test data.
    \item All three kernels are suffering from relatively low L2 cache hit rate, while the Mass and Solver kernels are suffering from a relatively low L1 cache rate.
\end{itemize}

\section{Conclusion}
\label{sec:conc}

This report concludes a series of papers on a grid-based structure preserving Landau collision operator with advanced numerical methods and a performance portable implementation in the PETSc numerical library \cite{Hirvijoki2016,AdamsHirvijokiKnepleyBrownIsaacMills2017,Adams2022a}.
This Landau solver supports multiple independent grids to efficiently resolve the domain of each species group, with multiple species per grid for species with like velocity profiles, and high-order accurate finite element discretizations with static, fully unstructured and block-structured, adaptive mesh refinement.
A new batch solver has been introduced and experiments with a well optimized code on an NVIDIA A100 and an AMD MI250X node is presented.
A new anisotropic relaxation test is presented that shows good agreement with analytical models and other published results.

The entire implicit time advance, after an initial setup phase, is written in the Kokkos programming language, and good hardware utilization is demonstrate, especially given the relative complexity of the kernel.
We observe 57\% FP64 pipe utilization (i.e., theoretical peak flop rate) on the NVIDIA A100 in the main computational kernel, the $2V$ Jacobian matrix construction, and comparable overall performance on the AMD MI250X. 
Many expressions in the Landau kernel have three or four multiply operands resulting in only $62\%$ of the flops being in fused multiple add instructions on the A100.
Thus, the 57\% theoretical peak is about $70\%$ of the theoretical peak for this algorithm on this hardware. 

Both the Landau time advance and batched linear solvers are publicly available in the PETSc (Portable Extensible Toolkit for Scientific Computing) numerical library, and the two performance test codes are examples in PETSc (Appendix \ref{sec:ad}).
%An anisotropic temperature relaxation verification test demonstrates good agreement with analytical results. 
%In related a future work, we develop this system  within the structure-preserving metriplectic formalism \cite{Kraus2017}, in which the symplectic and metric parts are evolved separately with structure-preserving time integrators that are unique to each system.
This grid-based method complements new full $3V$ particle-based Landau operators currently under development \cite{Zonta2021,Hirvijoki2021,PusztayKnepleyAdams2022}, and this grid-based method can be used with PIC codes with new conservative translation operators between particle and grid representations of distribution functions \cite{PusztayKnepleyAdams2022}.
Future work involves understanding the entropy generation of this translation method, which is a type of \textit{coarse-graining} used to add numerical entropy \cite{brunner1999collisional,Chen2007-CG,Vernay2012-cg}.

\section*{Acknowledgments}
This work was supported in part by the U.S. Department of Energy,
Office of Science, Office of Advanced Scientific Computing Research,
Scientific Discovery through Advanced Computing (SciDAC) program
through the FASTMath Institute under Contract No. DE-AC02-05CH11231 at
Lawrence Berkeley National Laboratory.

\appendix

\section{Artifact description and reproducibility}
\label{sec:ad}

PETSc output files with all data, provenance information, and reproducibility instructions for all tables and plots can be obtained from \path{git@gitlab.com:markadams4/batch_paper_data.git}.
This includes the python scrips that generates the plots and run scripts, makefiles and PETSc resource files used to generate the data.
The \path{landau_throughput_nsight} directory has throughput and NVIDIA hardware utilization data, and \path{landau_anisotropic_shift} has 
verification test data and AMD hardware utilization data.
The exact PETSc versions (SHA1) are in the data files, with the provenance data, and any PETSc version from v3.19 should suffice to reproduce this data. 
The two driver codes are in \path{src/ts/utils/dmplexlandau/tutorials}, ex1.c (anisotropic temperature relaxation \S\ref{sec:perfanisotropic}) and ex2.c (throughput studies in \S\ref{sec:throughput_perf}).
%PETSc must be configure for explicitly for the $3V$ tests and some scrips append the dimension to the executable name.

\section{NRL Plasma Formulary isotropization rates}
\label{sec:nrl}

The NRL Plasma Formulary provides relaxation evolution equations for inter and intra-species thermalization \cite{Huba2013}, according to 
$$ \frac{dT_\perp}{dt} = -\frac{1}{2}\frac{dT_\parallel}{dt} = -v_T^\alpha(T_\perp - T_\parallel),$$
where if $A \equiv T_\perp / T_\parallel - 1 > 0$,
$$ v_T^\alpha = \frac{2 \sqrt{\pi} e^2 n_\alpha \ln\Lambda_{\alpha\alpha}}{m_\alpha^{1/2}(kT_\parallel)^{3/2}}A^{-2}\left[ -3 + \left( A+3\right)\frac{tan^{-1}\left(A^{1/2}\right)}{A^{1/2}} \right].$$
The two species thermal equilibrium evolution equation is 
$$ \frac{dT_\alpha}{dt} = \bar{v} \left(T_\alpha - T_\beta\right),$$
 and 
$$ \bar{v} = 1.8 \times 10^{-19} \frac{\sqrt{m_e m_i} n_0 \ln\Lambda_{ei}}{\left( m_eT_i + m_iT_e\right)^{3/2}} sec^{-1}.$$

\section{Additional anisotropic relaxation test data}
\label{ap:details}

\begin{figure}[htbp]
\begin{center}
    \begin{subfigure}{0.49\linewidth} \centering
        \includegraphics[width=\linewidth]{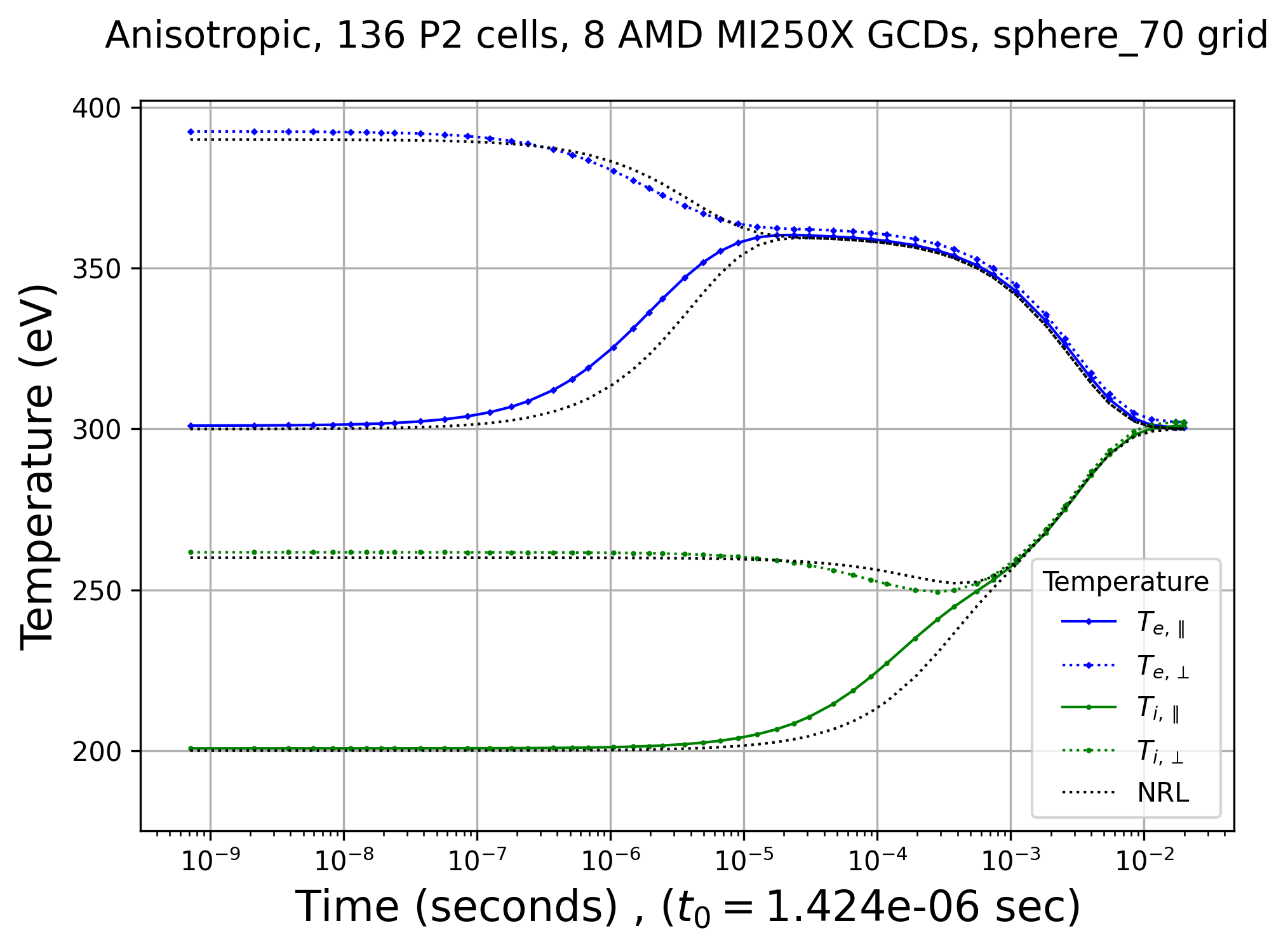}
    \end{subfigure}
    \begin{subfigure}{0.49\linewidth} \centering
        \includegraphics[width=\linewidth]{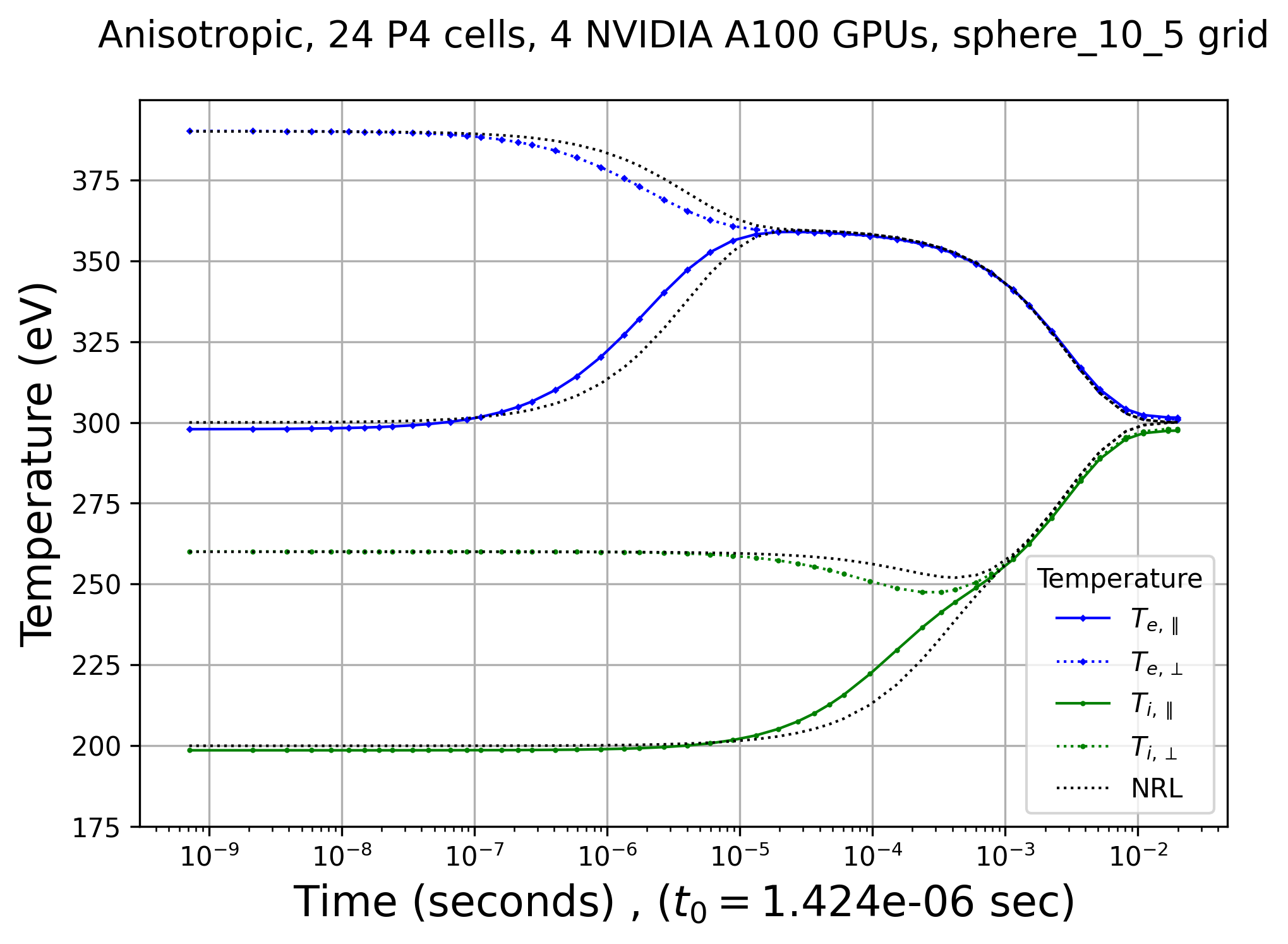}
    \end{subfigure}
\caption{Anisotropic relaxation test data from the $P2$ element case (left) and the $P4$ elements (right), plotted with analytical NRL model data}
\label{fig:temperature-history-extra}
\end{center}
\end{figure}

\begin{figure}[htbp]
\begin{center}
    \begin{subfigure}{0.49\linewidth} \centering
        \includegraphics[width=\linewidth]{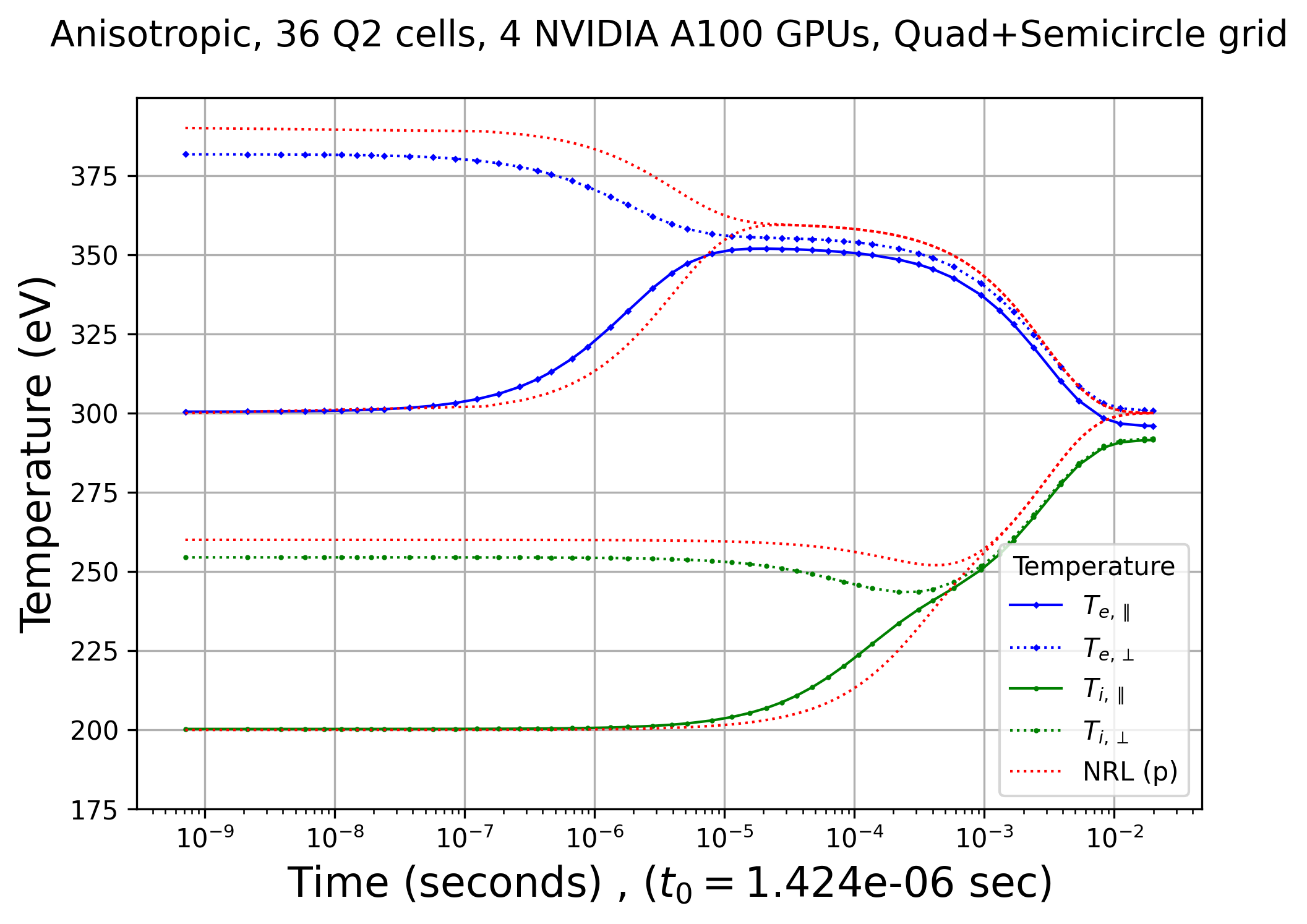}
    \end{subfigure}
    \begin{subfigure}{0.49\linewidth} \centering
        \includegraphics[width=\linewidth]{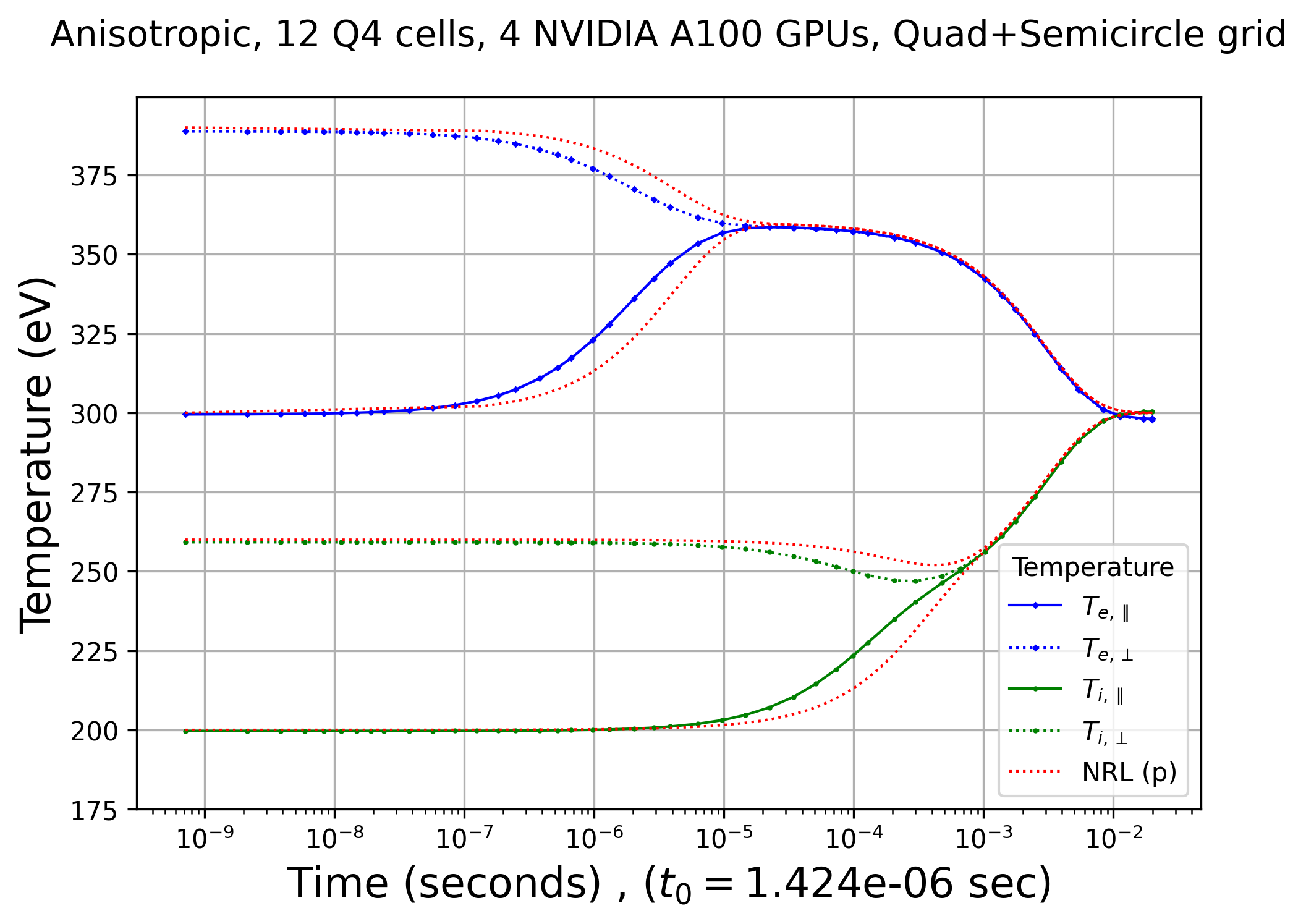}
    \end{subfigure}
\caption{Temperature vs. time with unstructured quadrilateral meshes and $Q2$ element case (left) and the $Q4$ elements (right), plotted with analytical NRL model data}
\label{fig:temperature-history-quad-extra}
\end{center}
\end{figure}

Figure \ref{fig:temperature-history-extra} shows anisotropic relaxation test data from the non-shifted Maxwellian case with $P2$ and $P4$ elements, where inaccuracies are visible.
Figure \ref{fig:temperature-history-quad-extra} shows anisotropic relaxation test data on the semicircle quadrilateral mesh where the poor quality of the $Q2$ mesh is clearly visible.

\bibliographystyle{siamplain}
\bibliography{cas-refs,the}





% \end{thebibliography}
\end{document}